\documentclass{aa}  

\usepackage{graphicx}
\usepackage{txfonts}
\usepackage{lipsum}
\usepackage{lscape}             
\usepackage{placeins}           

\usepackage{amsmath,amsfonts,amssymb}
\usepackage{textcomp,gensymb}
\usepackage[T1]{fontenc}
\usepackage{aecompl}
\usepackage{ulem}
\usepackage{aas_macros}
\usepackage[dvipsnames]{xcolor}
\usepackage{color}
\usepackage{times}
\usepackage{stfloats}
\usepackage[utf8]{inputenc}
\usepackage{hyperref}
\usepackage{multirow}
\usepackage{natbib}
\usepackage{bm}

\newcommand{\mdy}[2]{#2}                           

\newcommand{\mdytwo}[2]{#2}                           

\begin{document}
   \title{Density-driven Dust Growth in Binary Systems: \\ Inhibition of dust settling and growth in circumbinary discs 
   }
   \titlerunning{Dust Growth in Binary Systems: \\ Inhibition of dust settling and growth in circumbinary discs}
   \author{Antoine Alaguero\inst{1}
        \and
          Nicol\'as Cuello\inst{1}
        \and
          Jean-Fran\c cois Gonzalez\inst{2}
        \and
          Daniel J. Price\inst{3}
        \and
          Maxime Lombart\inst{4}
        \and
          Jeremy L. Smallwood\inst{5}
        \and
          Philippe Thébault\inst{6}
        }
   \institute{{Univ. Grenoble Alpes, CNRS, IPAG, 38000 Grenoble, France}\\
              \email{antoine.alaguero@univ-grenoble-alpes.fr}  
         \and
            {Université Lyon 1, ENS de Lyon, CNRS, CRAL, UMR 5574, Lyon, France} 
         \and
             {School of Physics and Astronomy, Monash University, Vic 3800, Australia}  
        \and
            {Université Paris-Saclay, Université Paris Cité, CEA, CNRS, AIM, 91191 Gif-sur-Yvette, France} 
        \and
            {Homer L. Dodge Department of Physics and Astronomy, The University of Oklahoma, Norman, OK 73019, USA} 
        \and
            {LIRA, Observatoire de Paris, Universite PSL ,5 Place Jules Janssen, 92195 Meudon, France} 
}

   \date{Received xxx; Accepted xxx}

 
  \abstract
   {Stellar multiplicity alters the density structure of protoplanetary discs and thereby the initial conditions for planet formation. Yet, the interplay between companion-disc interactions and dust growth remains poorly understood.}
   {The goal of this work is to investigate to what extent the density structure of a disc undergoing tidal interactions with a companion star promotes or inhibits the growth of dust grains.}
   {We perform a set of hydrodynamical simulations of protoplanetary discs orbiting one or both stars of a binary, including dust growth and fragmentation. We explore a range of companion orbits and compare the results with a single-star reference case.}
   {We find that dust growth is mainly driven by local accumulations of dust. In circumbinary discs, the maximum grain size is up to five times smaller than in isolated discs. This result likely originates from the perturbations \mdy{of}{caused by} the inner binary, which prevent dust grains \mdy{to properly settle and drift}{from properly settling and drifting}. As a consequence, the conditions required to trigger strong clumping driven by the streaming instability are difficult to achieve.
   In contrast, circumstellar discs in binary systems exhibit grain sizes similar to those in isolated discs, leading to comparable conditions for strong clumping by the streaming instability. }
   {Planet formation through core accretion seems challenging in circumbinary discs harbouring binaries larger than a few au, suggesting that circumbinary planets observed near the dynamical stability limit did not form \textit{in situ}. Conversely, perturbations from external companions only marginally affect density-driven dust growth compared to isolated systems.}


   \maketitle
   \nolinenumbers

\section{Introduction}
\label{sec:intro}

Sub-structures in protoplanetary discs are now routinely detected and characterised \citep{Andrews+2018,Benisty+2023}. High-resolution imaging shows that these features are widespread across a variety of stellar and disc environments, suggesting that they are common outcomes of disc evolution \citep{Garufi+2024,Vioque+2025-AGEPRO}. Sub-structures strongly influence the distribution and dynamics of gas and solids, shaping the earliest stages of planet formation in the process \citep{Bae+2023}. Understanding their origin, prevalence, and evolution is therefore essential to understanding planet formation. To this day, proposed formation pathways of sub-structures include planet-disc interactions \citep{Paardekooper+2023}, disc-environment interactions \citep[e.g.][]{Calcino+2025}, physical processes within discs \citep[e.g.][]{Gonzalez+2017,Riols+2020}, radiative interactions \citep[e.g.][]{MontesinosCuello2018}, or interactions with companion stars \citep[e.g.][]{Gonzalez+2020,Cuello+2023}.

In the last case, sub-structures arise from tidal forces exerted by companion stars on the disc. These gravitational perturbations result in a wide range of morphological features. First, depending on the location of the companion, tidal truncation either carves an inner cavity in the disc, or reduces its radial extent from the outside \citep{ArtymowiczLubow1994}. Second, spiral density waves launched at the Lindblad resonances propagate through the disc \citep{Rafikov2002}, leaving detectable kinematic signatures \citep[e.g.][]{Alaguero+2024}. Third, horseshoe-shaped overdensities may develop in circumbinary discs \citep{Ragusa+2017}. The properties of all these features depend sensitively on the orbital configuration of the system \citep{Calcino+2023,Cuello+2025}. Dust and gas in circumstellar and circumbinary discs therefore evolve in a dynamical environment distinct from that of discs around single stars.

In the core accretion scenario \citep{Lissauer1993}, sub-structures strongly affect dust growth. 
In multiple-star systems, they can efficiently promote the accumulation of solids. For example, binary-disc interactions result in a natural pressure bump that traps dust grains at the inner edge of circumbinary discs \citep{Cazzoletti+2017}. In circumstellar discs in binary systems, spiral arms and interactions with an external star can also concentrate dust \citep[][\mdy{}{seeing binary systems as repeated flybys}]{Prasad+2025,Su+2026}. Additionally, spiral arms and horseshoe asymmetries form regions of enhanced gas density. At fixed dust density, higher gas density increases the dust-gas aerodynamic coupling, leading to lower grain impact velocities and thereby promoting their growth \citep{Birnstiel+2016}. Multiple-star systems may therefore help overcome the barriers posed by radial drift and fragmentation \citep{Adachi+1976,Weidenschilling1977,BlumWurm2008}.

Yet these environments can also hinder sustained dust growth.
Repeated perturbations by a stellar companion increase collisional velocities between dust grains, making fragmentation more likely. Close to the inner edge of circumbinary discs, collisions amongst centimetre-sized particles can reach velocities of $\sim40$~m~s$^{-1}$ \citep{Pierens+2021}. Comparable collisional velocities may occur in circumstellar discs perturbed by external companions, similarly to discs with embedded planets \citep{Eriksson+2025}. 

Moreover, dust evolution is not a passive response to disc dynamics: changes in grain size distribution and opacity can modify radiative cooling and dust-gas coupling, which may in turn generate new sub-structures or trigger planet-forming instabilities \citep{Gonzalez+2017,Ho+2024,Lee+2025}.
The relationship between dust evolution, disc morphology, and disc kinematics is thus inherently interdependent and non-linear.

Dust growth has been studied extensively in discs around single stars, but remains poorly quantified in multiple-star systems despite their prevalence at young stellar ages \citep{Chen+2013,Offner+2023}. The dynamical environments of multiple systems provide a natural laboratory for examining dust evolution under strong perturbations. 
Tidal perturbations from stellar companions lead to destructive grain collisions, a reduction of the available material for planet formation, and susceptibility to warping, misalignment, and disc breaking \citep{Manara+2019,MarzariThebault2019,Zurlo+2023,Rabago+2024}.
Assessing how planet formation occurs in such severe conditions represents the first step toward the understanding of the exoplanet population detected in multiple-star systems \citep{ThebaultBonanni2025}. Previous studies have either focused on gas dynamics or adopted simplified dust prescriptions neglecting the feedback between particle growth, disc structure, and disc dynamics.
In this series of papers, we investigate dust growth and fragmentation in binary systems using three-dimensional hydrodynamical simulations including dust grains of evolving size. We simulate both circumstellar and circumbinary discs. In this first paper, we examine how the density structure of discs in binary systems promotes or inhibits dust coagulation, and how evolving dust populations may in turn modify these structures. Our goal is not to predict precise grain sizes in multiple systems, but to isolate the effects of density perturbations on dust growth. A companion paper (Alaguero et al., in prep.) will extend this analysis by examining how disc kinematics affects the relative velocities of dust grains. Together, these studies will enable robust predictions of the grain sizes reached in structured discs.  
Our broader aim is to identify regions of protoplanetary discs in binary systems that are favourable to planet formation and to clarify the interplay between binary-driven disc dynamics and dust evolution during the earliest stages of planet formation.

This first paper is organised as follows: we describe our methods and the simulation setup in Section \ref{sec:methods}, then we analyse our simulations in Section \ref{sec:results}, before discussing our results in Section \ref{sec:discussion} and concluding in Section \ref{sec:conclusion}.

\section{Methods}
\label{sec:methods}

We performed 3D hydrodynamical simulations of protoplanetary discs in binary systems. We used the smoothed particle hydrodynamics \citep[SPH, e.g.][]{GingoldMonaghan977,Lucy1977} code {\sc Phantom} \citep{Price+2018-phantom}. The simulations included 
$1.5\times10^6$ gas+dust particles following a dust-as-a-mixture scheme \citep{LaibePrice2014c,LaibePrice2014a,LaibePrice2014b,PriceLaibe2015,Ballabio+2018}. The purpose of this work is to study the ability of small grains ($\sim60$ $\mu$m) to grow. These grains are tightly coupled to the gas and validate the terminal velocity approximation \citep{YoudinGoodman2005}, motivating the use of a dust-as-a-mixture scheme (see Appendix \ref{app : St}). 

The size of dust grains was allowed to evolve following the monodisperse implementation described in \cite{Vericel+2021}: each SPH particle represented a swarm of identical spherical grains colliding with each other and evolving in size as a function of these collisions. We assumed grains to be compact and spherical, and gas turbulence to be the main driver of collisions between dust grains\footnote{other contributions, coming for example from differential radial drift, vertical settling, or brownian motions are neglected. First, these differential motions imply collisions between grains of different sizes, which cannot be treated in the monodisperse approximation. Second, for grains larger than a few tens of micrometers, gas turbulence dominates over the other contributions \citep{Birnstiel+2016}.}. In this context, the work of \cite{StepinskiValageas1997} provides analytical expressions to the collisional terminal velocity between dust grains, \mdy{noted}{denoted} $V_{\text{rel}}$,
\begin{eqnarray}
    \label{eq:vrel}
    V_{\text{rel}} &=& \sqrt{2} V_{\mathrm{t}} \frac{\sqrt{\mathrm{Sc} - 1}}{\mathrm{Sc}} \,,\\
    V_{\mathrm{t}} &=& c_{\mathrm{s}}\sqrt{\sqrt{2}\alpha_{\text{SS}}\mathrm{Ro}} \,,\\
    \mathrm{Sc} &=& (1 + \mathrm{St}) \sqrt{1 + \frac{|\Delta \bm{v}|^2}{V_{\mathrm{t}}^2}} \,,
\end{eqnarray}
where $V_{\mathrm{t}}$ is the gas turbulent velocity prescribed as a function of the Rossby number $\mathrm{Ro}=3$ and of the $\alpha_{\text{SS}}$ viscosity parameter \citep{ShakuraSunyaev1973}. $\mathrm{Sc}$ is the Schmidt number, $c_{\mathrm{s}}$ is the sound speed, and $\Delta \bm{v}$ is the differential velocity between dust and gas. Finally, the Stokes number is \mdy{noted}{denoted} $\mathrm{St}$ and defined as a function of the dust stopping time due to gas drag, \mdy{noted}{denoted} $t_{\mathrm{s}}$, and of the Keplerian frequency, \mdy{noted}{denoted} $\Omega_{\mathrm{K}}$,
\begin{equation}
    \label{eq:omega_k}
    \mathrm{St} = t_{\mathrm{s}}\, \Omega_{\mathrm{K}} \,,
\end{equation}
The expression and calculation of $t_{\mathrm{s}}$ depend on the drag regime \citep[e.g.][]{LaibePrice2012b}. Our simulations took place in the Epstein regime, for which the expression of the stopping time is \citep{Epstein1924}
\begin{eqnarray}
    \label{eq:t_s}
    t_{\mathrm{s}} &=& \frac{\rho_{\mathrm{s}} s}{(\rho_{\mathrm{d}}+\rho_{\mathrm{g}})\,c_{\mathrm{s}}\,f}  \sqrt{\frac{\pi\gamma}{8}} \,,\\
    f &=& \sqrt{1+\frac{9\pi}{128}\frac{|\Delta \bm{v}|^2}{c_{\mathrm{s}}^2}} \,,
\end{eqnarray}
with $\rho_{\mathrm{s}}$ being the grain intrinsic density set to $3.0\times10^3$~kg~m$^{-3}$, $s$ the grain size, $\rho_{\mathrm{d}}$ the dust density, $\rho_{\mathrm{g}}$ the gas density, and $\gamma$ the adiabatic index set to $1$.

In this formalism, $V_{\text{rel}}$ assumes a Keplerian rotation profile and is therefore insensitive to the non-Keplerian disc dynamics induced by the gravitational perturbation of the companion star. We plan to address this contribution in a follow-up paper. The present formulation of $V_{\text{rel}}$, however, is sufficient to examine how the morphological changes induced by stellar companions in the disc structure affect dust growth, which is the focus of this paper. $V_{\text{rel}}$ was compared to a fragmentation velocity, \mdy{noted}{denoted} $V_{\text{frag}}$, in order to discriminate between the growth and fragmentation regimes. We chose $V_{\text{frag}}=15$~m~s$^{-1}$, which is consistent with grains coated with water ices \citep{ShimakiArakawa2012,GundlachBlum2015} and with previous studies \citep{GarciaGonzalez2020,Vericel+2021,Michoulier+2024}. 
Grain growth was treated as perfect coagulation, and fragmentation was smoothed to consider a progressive loss of mass with increasing $V_{\text{rel}}$ (\citealt{KobayashiTanaka2010}, see also equation 15 of \citealt[][]{Vericel+2021}).

In each simulation, we initialised a disc orbiting either one or two stars of a binary. The stars were represented by sink particles of corresponding mass \citep{Bate+1995}. We varied the eccentricity of the pairs of stars between simulations but we kept their semi-major axis constant. We considered four orbital setups, in which the disc was either a circumbinary disc or a circumstellar disc in a binary system, and the orbital eccentricity was either taken as $e_{\mathrm{bin}}=0$ or $e_{\mathrm{bin}}=0.5$. Additionally, we ran a control simulation with a disc orbiting a single star. Hereafter, these different orbital setups will be \mdy{noted}{denoted} as follows:
\begin{itemize}
    \item \textit{IB0}: a circumbinary disc orbiting a binary star with $e_{\mathrm{bin}}=0$,
    \item \textit{IB5}: a circumbinary disc orbiting a binary star with $e_{\mathrm{bin}}=0.5$,
    \item \textit{OB0}: a circumstellar disc orbiting one of the \mdy{star}{stars} of a binary star with $e_{\mathrm{bin}}=0$,
    \item \textit{OB5}: a circumstellar disc orbiting one of the \mdy{star}{stars} of a binary star with $e_{\mathrm{bin}}=0.5$,
    \item \textit{ref}: a disc orbiting a single star.
\end{itemize}
The disc orbited a central mass of $1$~M$_{\odot}$. In the \textit{IB} simulations, the binary had a mass ratio\footnote{defined as $q=\frac{M_1}{M_1+M_2}$, with $M_1$ and $M_2$ the masses of the primary and secondary, respectively.} of $q=0.75$ and a semi-major axis of $a_{\mathrm{bin}}=5$~au.
These parameters were chosen to represent circumbinary discs with observed central cavities that harbour binaries of a few au, which are commonly observed and a natural outcome of star formation processes \citep{Czekala+2019,Elsender+2023}. In the \textit{OB} simulations, the perturber had a mass of $0.1$~M$_{\odot}$ and a semi-major axis of $a_{\mathrm{bin}}=100$~au. These parameters were chosen to represent a low-mass stellar companion on a wide orbit, a configuration commonly observed in young multiple systems \citep{Tobin+2022}. At this separation \mdy{}{and given an initial disc outer radius of $50$~au}, the companion exerts significant tidal perturbations on the disc without completely disrupting it. This regime allows the companion to generate observable density structures while preserving a substantial disc reservoir in which dust growth and planet formation can be investigated.

The sink particles all had an accretion radius of $1$~au, except for the perturber in the \textit{OB} simulations which had an accretion radius of $5$~au. We chose these values to mitigate the formation of circumsecondary discs, which are computationally demanding to resolve and the study of which is beyond the scope of this work. 
We initialised the binary and the disc in a coplanar configuration. Table \ref{table:varied_params_dg} and Table \ref{table:fixed_params_dg} summarise the parameters that were varied or fixed across the simulations, respectively.

The initial disc followed an exponentially tapered power-law in surface density with a tapered inner edge between the initial inner radius $R_{0,\mathrm{in}}$ and the initial outer radius $R_{0,\mathrm{out}}$, written as
\begin{equation}
\label{eq:powerlaw_exp_disc_sigma}
     \Sigma(r) = \Sigma_0 \left(\frac{r}{R_0}\right)^{-p} \exp\left({-\frac{r}{R_{\mathrm{c}}}}\right)^{2-p} \left(1-\sqrt{\frac{R_{0,\mathrm{in}}}{r}}\right)\, , 
\end{equation}
with $R_0$ being a radius of reference, $\Sigma_0$ the surface density at that radius, $R_{\mathrm{c}}$ the exponential cut-off radius, and $p$ the power-law index for the density. We normalised the surface density for the disc to have a total gas mass of $M_{\mathrm{g}}=0.05$~M$_{\odot}$. We chose a value of $R_{0,\mathrm{in}}=2a_{\mathrm{bin}}$ in \textit{IB0} and \textit{IB5} in order to ease the disc relaxation.
We assumed the equation of state to be locally isothermal around the central sink(s). We fixed the temperature to a power-law radial profile with an exponent of $q=0.5$ normalised for the aspect ratio of the disc to be $\frac{H}{R}\vert _{R_0}=0.05$, with $H$ the gas scale height. We tuned the SPH viscosity to achieve a Shakura-Sunyaev viscosity parameter of $\alpha_{\text{SS}}=5\times10^{-3}$ \citep{LodatoPrice2010}.

The simulations were first evolved without dust for \mdy{$130$}{$9$} periods of the disc at $R_{0,\mathrm{out}}$, \mdy{noted}{denoted} $P_{\text{disc}}$, in order for the disc to relax from the initial conditions and prevent numerical artifacts in the dust distribution. The dust-free simulations were integrated until achieving negligible changes in disc outer radius, that we note $R_{\mathrm{out}}$, over time. After \mdy{$130$}{$9$}~$P_{\text{disc}}$, all the discs showed changes\footnote{for that measurement $R_{\mathrm{out}}$ was taken as the radius encircling $63.8\%$ of the disc mass, similarly to \cite{Bate2018}.} in $R_{\mathrm{out}}$ below $0.2$~au over the last \mdy{$10$}{}~$P_{\text{disc}}$.
For \textit{OB0} and \textit{OB5}, we introduced the external star only after the disc had relaxed. This prevented the formation of sub-structures, such as large spiral arms, before adding dust. In circumbinary discs, however, introducing a star after the relaxation phase would strongly disturb the inner disc and require an additional relaxation period of $\sim200$~$P_{\text{bin}}$ to establish the central cavity. For this reason, the \textit{IB0} and \textit{IB5} discs were relaxed from the outset as circumbinary systems.

After the relaxation period, we added dust following the dust-as-a-mixture scheme. The spatial distribution of dust was initially assumed identical to that of the gas after relaxation and with a dust-to-gas mass ratio of $\delta_{\mathrm{dg}}=0.01$. To save computational time, we removed particles outside a radius $R_{\text{clear,out}}=2a_{\mathrm{bin}}$ in \textit{OB0} and \textit{OB5}, and outside a radius $R_{\text{clear,out}}=4$ $R_{0,\mathrm{out}}$ in \textit{IB0} and \textit{IB5}. In \textit{IB0} and \textit{IB5}, we also deleted particles inside a radius $R_{\text{clear,in}}=1.5a_{\mathrm{bin}}$. 

After the relaxation, dust grains started with a size of $s_0=60$~$\mu$m. We floored the grain size to $s_{\text{min}}=50$~$\mu$m to avoid possible numerical errors produced by very small grains\footnote{Time-step errors can lead to a bad estimation of the time derivative of grain mass. If grains are fragmenting, overshooting the time derivative result in a negative grain mass. A minimal grain size is defined to prevent this effect.}. \cite{Vericel+2021} have shown that grains starting with $s_0<50$~$\mu$m quickly reach similar sizes \mdy{that}{to} grains starting with $s_0\sim50$~$\mu$m, proving that the choice of $s_0$ is not critical as long as it remains small compared to the final grain size. Since the interest of this work is to highlight the regions favourable to growth, knowing the full grain size distribution is also not necessary. For this reason, we used a monodisperse model.

The simulations evolved for \mdy{$1000$}{$83$}~$P_{\text{disc}}$ after the initial relaxation time. \mdy{It corresponded}{This duration corresponds} to $29$~kyr, which is approximately equivalent to $2635$~$P_{\text{bin}}$ and $29$~$P_{\text{bin}}$ in the \textit{IB} and \textit{OB} configurations respectively.

\begin{table}[h!]
\caption{Variable initial parameters of the dust growth hydrodynamical simulations.} 
\centering
 \begin{tabular}{c c c c c c} 
 & \textit{IB0} & \textit{IB5} & \textit{OB0} & \textit{OB5} & \textit{ref} \\
 \hline \hline

 $M_{1}$ (M$_{\odot}$) & $0.75$ & $0.75$ & $1$ & $1$ & $1$ \\
 \hline
 $M_{2}$ (M$_{\odot}$) & $0.25$ & $0.25$ & $0.1$ & $0.1$ & / \\
 \hline
 $a_{\mathrm{bin}}$ (au) & $5$ & $5$ & $100$ & $100$ & / \\
 \hline
 $e_{\mathrm{bin}}$  & $0$ & $0.5$ & $0$ & $0.5$ & / \\
 \hline
 $R_{0,\mathrm{in}}$ (au)  & $10$ & $10$ & $3$ & $3$ & $3$ \\
 \hline
\end{tabular}
\label{table:varied_params_dg}
\end{table}

\begin{table}[h!]
\caption{Fixed initial parameters of the dust growth hydrodynamical simulations.} 
\centering
 \begin{tabular}{c c} 
   Parameter & Value \\ 
 \hline \hline
 $M_{\mathrm{g}}$ (M$_{\odot}$) & $5.0\times10^{-2}$\\
 \hline
 $M_{\mathrm{d}}$ (M$_{\odot}$) & $5.0\times10^{-4}$\\
 \hline
 $R_{\mathrm{c}}$ (au) & $40$\\
 \hline
 $R_{0,\mathrm{out}}$ (au) & $50$\\
 \hline
 $R_0$ (au) & $20$\\
 \hline
 $\frac{H}{R}\vert _{R_0}$ & $5.0\times10^{-4}$ \\ 
 \hline
 $p$ & $1$\\
 \hline
 $q$ & $0.5$ \\
 \hline
 $\alpha_{\mathrm{SS}}$ & $5.0\times10^{-3}$ \\
 \hline
 $V_{\text{frag}}$ (m s$^{-1}$) & $15$ \\
 \hline
 $s_{0}$ ($\mu$m) & $60$ \\
 \hline
 $s_{\mathrm{min}}$ ($\mu$m) & $50$ \\
 \hline
 $\rho_{\mathrm{s}}$ (kg m$^{-3}$) & $3.0\times10^{3}$ \\
 \hline
 $\delta_{\mathrm{dg}}$ & $10^{-2}$ \\
 \hline
\end{tabular}
\tablefoot{All the systems were initialised in a coplanar configuration.}
\label{table:fixed_params_dg}
\end{table}

\section{Results}
\label{sec:results}

\subsection{Density}
\label{subsec:density}

\begin{figure*}[ht]
\centering
\begin{center}
    \includegraphics[width=\textwidth, trim={1cm 4cm 1cm 2cm},clip]{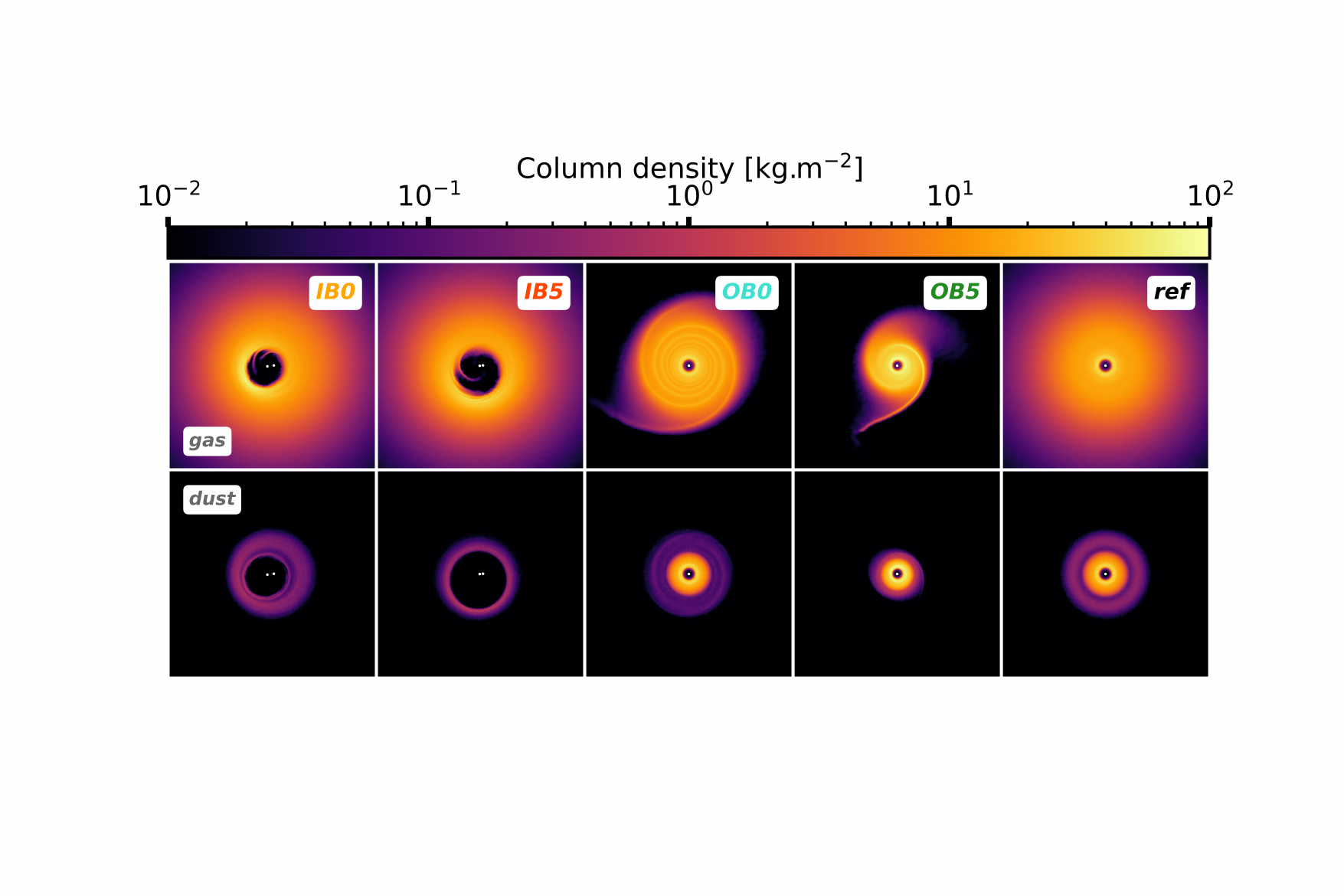}     
    \caption{
    Column density rendering of the simulations at $29$ kyr. The top row shows the gas column density and the bottom row the dust column density. From left to right, the columns correspond to \textit{IB0}, \textit{IB5}, \textit{OB0}, \textit{OB5}, and \textit{ref}, respectively. The dimensions of the boxes are $160$~au $\times$ $160$~au. }
    \label{fig:all_density}
\end{center}
\end{figure*}

\begin{figure*}[ht]
\centering
\begin{center}
    \includegraphics[width=\textwidth, trim={0cm 0cm 0cm 0cm},clip]{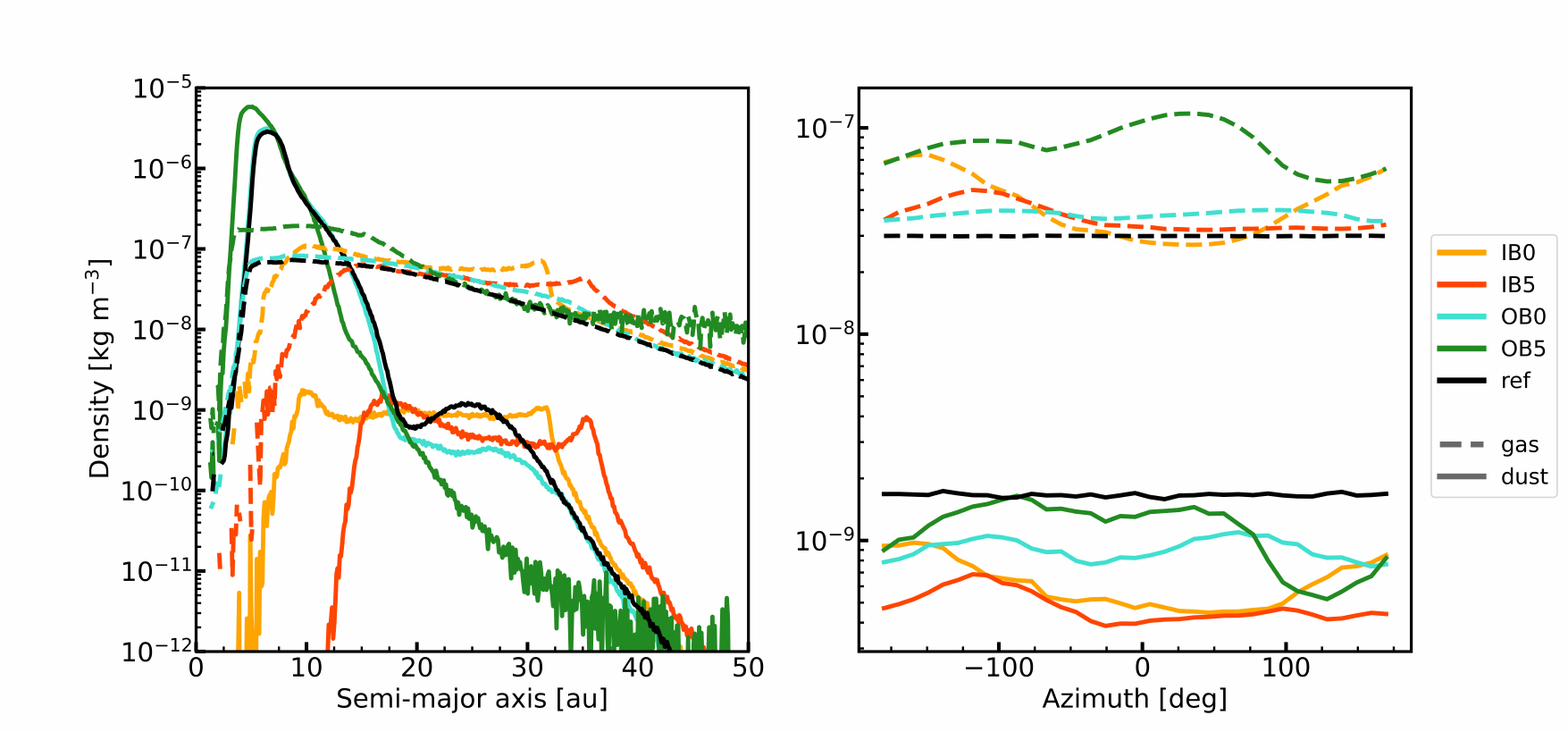}     
    \caption{
    \textbf{Left:} Radial profiles of the density at $29$ kyr in each simulation.  
    \textbf{Right:} Azimuthal profiles of the density at $29$ kyr in each simulation. The density was averaged on particles with a semi-major axis between $15$~au and $50$~au. In both plots the solid lines represent the dust and the dashed lines represent the gas.
    }
    \label{fig:density_radazprof}
\end{center}
\end{figure*}

Figure \ref{fig:all_density} shows the dust and gas column densities at the end of the simulations. Figure \ref{fig:density_radazprof} shows the corresponding density radial and azimuthal profiles of dust and gas. The azimuthal profiles were computed by averaging the density radially between $15$~au and $50$~au.

Circumbinary discs (\textit{IB0} and \textit{IB5}) exhibit an eccentric inner cavity, that becomes larger when the binary is eccentric. Dust is depleted at small radii ($\lesssim10$~au for \textit{IB0} and $\lesssim15$~au for \textit{IB5}), with density rising only farther out and forming a plateau beyond the inner disc edge. 
These discs also show an azimuthal asymmetry caused by a lump at the cavity edge (approximately between $-50\degree$ and $100\degree$ for \textit{IB0}, and between $-170\degree$ and $-50\degree$ for \textit{IB5}). Such a feature is expected from disc-binary interactions and is more prominent when the binary is circular \citep{Ataiee+2013,Ragusa+2017,Ragusa+2020,Mignon-Risse+2023}. The asymmetry is present in both gas and dust.
Near the cavity ($<25$~au), the gas disc is threaded by spiral arms. Gas penetrates closer in and is less sharply truncated, but its density in the inner regions remains lower than in the single-star case (see Appendix \ref{app : disc size} for details on disc size and eccentricity). Near the dust disc outer edge (at $31$~au for \textit{IB0} and $36$~au for \textit{IB5}), \mdy{the density profile shows}{both the gas and dust density profiles show} a bump. \mdy{}{To investigate the physical origin of these density bumps, we performed gas-only simulations for the \textit{IB0} and \textit{IB5} configurations. As shown in Appendix \ref{app : acc drift}, these simulations also exhibit a gas density bump at the outer disc edge. This implies that the gas pressure bump responsible for dust trapping is not generated by the dust back-reaction, unlike the self-induced dust traps described by \cite{Gonzalez+2017}. By contrast, the gas density profile of \textit{ref} is smooth. This suggests that the pressure bump observed in \textit{IB0} and \textit{IB5} is instead produced by the inner binary. More specifically, spiral density waves excited at the cavity edge propagate outwards, transporting angular momentum radially until they are damped. The resulting deposition of angular momentum may locally overcome the loss of angular momentum by viscous turbulence, leading to the accumulation of material and the formation of a pressure maximum. Future work will be key to constrain this process in more details.} 
\mdy{A similar feature appears in}{In} \textit{OB0} and \textit{ref}\mdy{}{, there is a local variation in the slope of the dust density profile} around $25$~au. \mdy{}{Yet, the gas density profile remains smooth in these simulations.} \mdy{This bump corresponds to an accumulation of drifting grains.}{We interpret this local dust density enhancement as the result of an accumulation of inward-drifting dust grains.} As grains drift inward at nearly constant size, they encounter progressively higher gas densities, which increases their aerodynamic coupling to the gas and reduces their drift velocity. \mdy{This process leads to local dust accumulations}{The resulting slowdown causes grains to accumulate locally, giving rise to the observed dust density enhancement} (see Appendix \ref{app : acc drift} for details).

The disc around an isolated star (\textit{ref}) and the circumstellar discs in binary systems (\textit{OB0} and \textit{OB5}) all show strong dust density peaks in the inner regions followed by a gradual decrease with radius. Circumstellar discs in binaries have smaller radii than the single-star disc due to tidal truncation by the companion (see Appendix \ref{app : disc size}). In particular in \textit{OB5} and compared to \textit{OB0}, the truncation is sharper and the dust density peaks higher because of the eccentricity of the binary \citep{MirandaLai2015}. This is because the dust disc size is also set by radial drift, which is accelerated in the presence of an external companion \citep{Manara+2019,Rota+2022}. 
The external companion triggers double-armed spirals in the gas disc at the Lindblad resonances, which then propagates outward. While these spirals are maintained over time in \textit{OB0}, they dissipate after each periastron passage in \textit{OB5}. Dust is slightly entrained by the spirals, which is why spirals form azimuthal asymmetries both in the dust and gas density profiles.
Across all cases, gas is more extended and smoother than dust.

\begin{figure*}[ht]
\centering
\begin{center}
    \includegraphics[width=\textwidth, trim={0cm 0cm 0cm 0cm},clip]{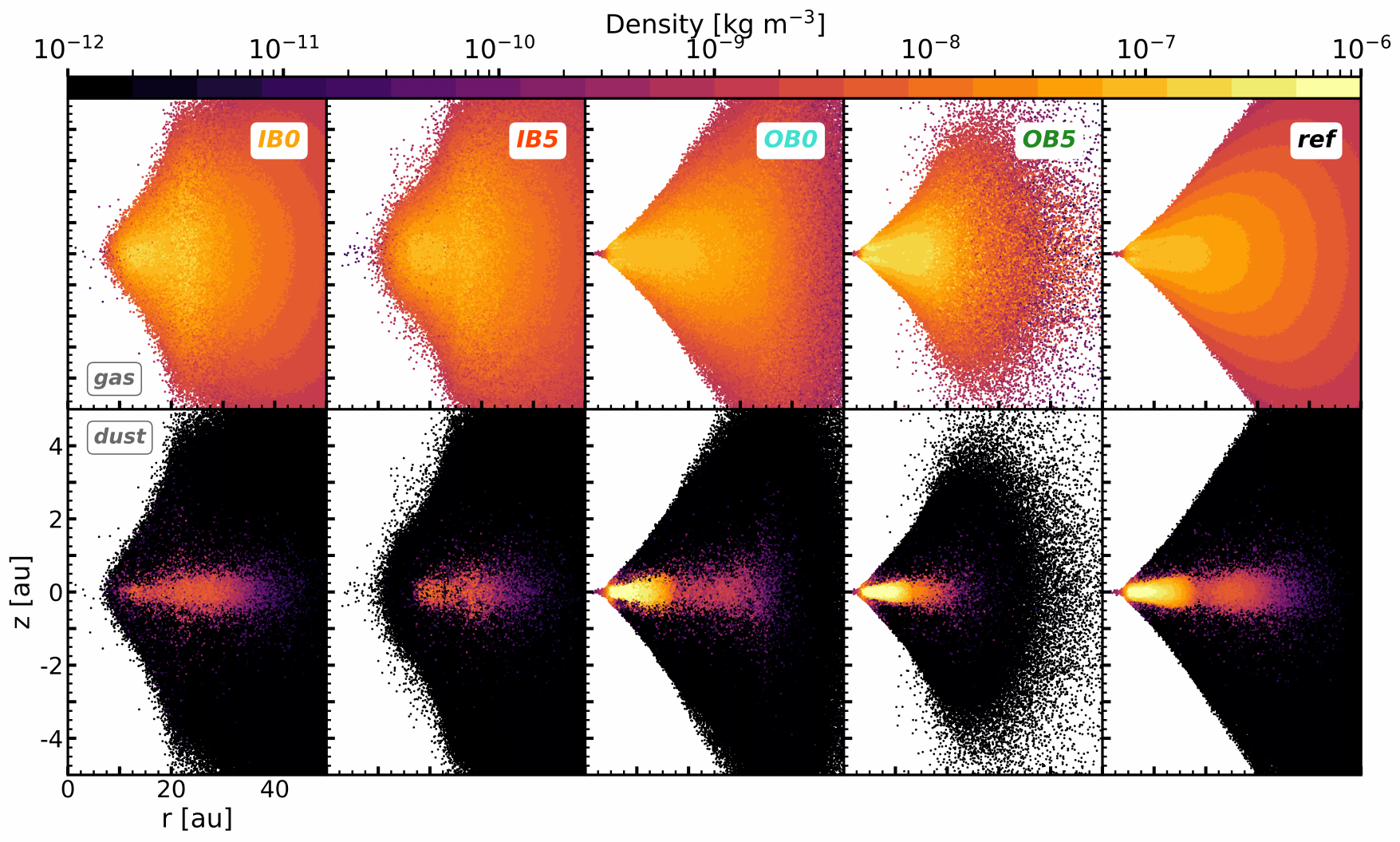}     
    \caption{Spatial distribution of the gas (top) and dust (bottom) density at $29$~kyr in each simulation. \mdy{}{The simulation name is indicated at the top right of each column. The panels display the full 3D data projected onto the plotting plane.}}
    \label{fig:rz_dustdensity}
\end{center}
\end{figure*}

Figure \ref{fig:rz_dustdensity} shows the gas and dust density distributions as a function of radius and altitude. The top row shows the gas density while the bottom row shows the dust density. The dust is concentrated close to the midplane, while the gas is more spread out. This discrepancy is naturally explained by the vertical settling of dust grains \citep[e.g.][]{GaraudLin2004,Barrierefouchet+2005}. Apart from this general dust-settling effect, stellar companions creates additional features in the gas and dust distributions.
The outer regions ($>20$~au) of the \textit{OB} discs are the most perturbed by the companion, yet there is almost only gas at these locations because of radial drift. In the inner regions ($\leq20$~au), the gas disc in \textit{OB5} is denser than in \textit{OB0} and \textit{ref}. This occurs because, at each periastron passage in \textit{OB5}, the disc material becomes eccentric and loses angular momentum, which drives inward migration \citep{Ostriker1994,Winter+2018}. Apart from that, \textit{OB0}, \textit{OB5}, and \textit{ref} have similar dust vertical distributions.
Compared to circumstellar discs, circumbinary discs are more vertically extended in the gas. This phenomenon has been previously characterised, and comes from the perturbation of the inner disc by the binary \citep{Papaloizou2005a,BarkerOgilvie2014,Pierens+2020}: the binary excites eccentricity in the inner disc, which becomes susceptible to a parametric instability that drives turbulent motions. As a result, the inner regions experience enhanced turbulence that hinders efficient radial drift and vertical settling of dust particles \citep[and Appendix \ref{app : inner cbd}]{Pierens+2021}. This produces a dust plateau between $10$ and $40$~au, where particles remain continuously stirred. Because drift and settling are suppressed, the resulting peak dust density is lower than in circumstellar discs.

\subsection{Dust grain size}
\label{subsec:grainsize}

\begin{figure*}[ht]
\centering
\begin{center}
    \includegraphics[width=\textwidth, trim={0cm 0cm 0cm 0cm},clip]{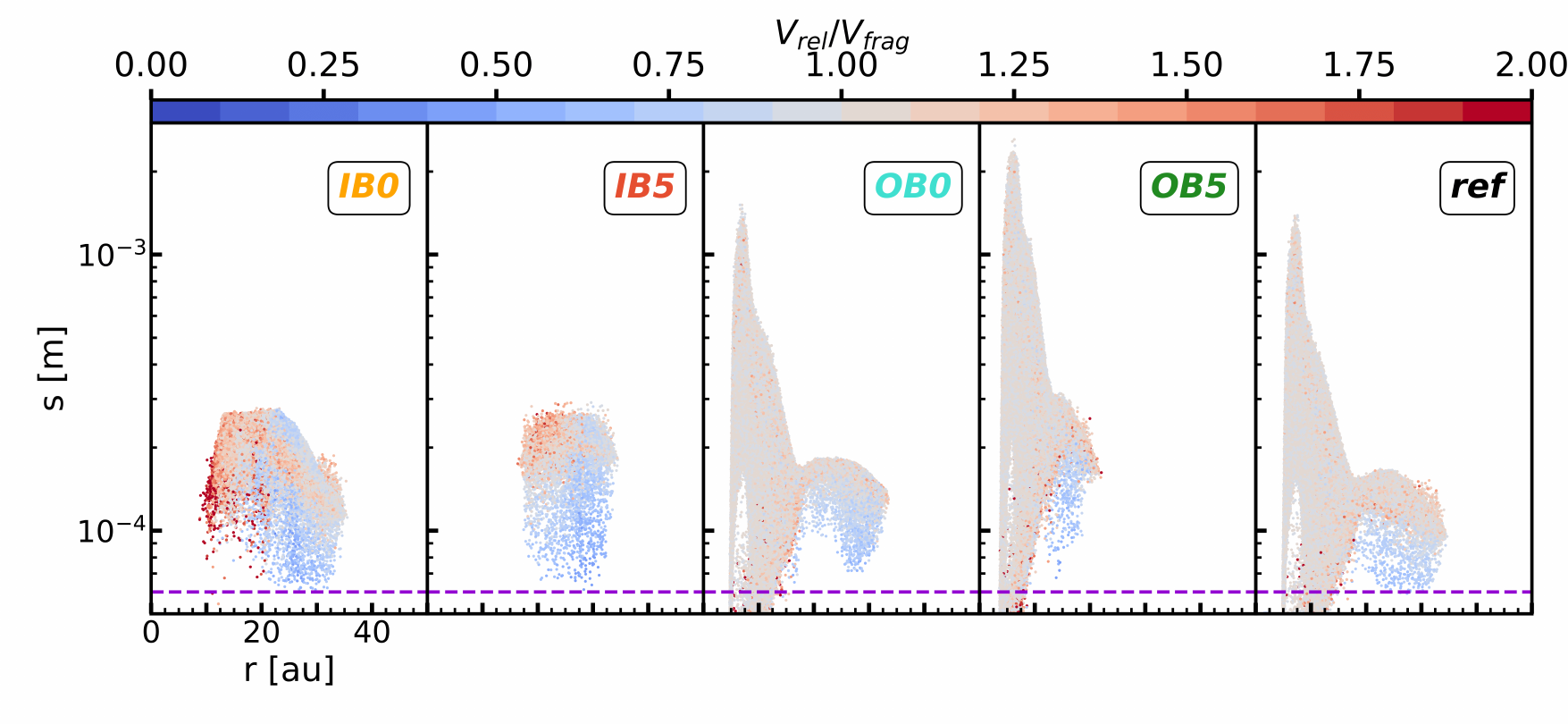}     
    \caption{Grain size represented by each SPH particle as a function of radius at $29$ kyr. The particles are coloured according to the $V_{\text{rel}}/V_{\text{frag}}$ ratio. The purple dashed line corresponds to the initial size of dust grains. The simulation name is indicated at the top right of each panel. \mdy{}{The panels display the full 3D data projected onto the plotting plane.} For a better visualisation, particles with a dust to gas ratio $\delta_{\mathrm{dg}}<0.01$ are filtered out.}
    \label{fig:rgrainsizevrel}
\end{center}
\end{figure*}

%
Figure \ref{fig:rgrainsizevrel} shows the grain size represented by SPH particles as a function of radius in each simulation. The purple dashed line represents the initial size of dust grains. We see that grains grew in size by approximately a factor $5$ in the \textit{IB} simulations. The grain size distribution becomes flat close to the maximum grain size between $~10$~au and $~28$~au for \textit{IB0} and between $~20$~au and $~35$~au for \textit{IB5}. This plateau is reminiscent of the plateau in dust density shown in Figure \ref{fig:density_radazprof} at the same location.
Concerning the simulations of circumstellar discs, the grains grew by more than a factor $25$ compared to the initial grain size. The grain size spatial distribution is peaked at the inner disc edge, where the density is maximum. \textit{OB0} and \textit{ref} present comparable distributions and peak levels of grain size. However, when compared to grains at the same radius in the latter two simulations, grains in \textit{OB5} are generally larger (except for grains beyond $~25$~au, where the disc in \textit{OB5} is truncated). The maximum grain size is larger by a factor $\sim1.7$ in \textit{OB5} compared to \textit{ref}. This, again, is consistent with the distribution of dust density.
The particles plotted in Figure \ref{fig:rgrainsizevrel} are coloured according to their $V_{\text{rel}}/V_{\text{frag}}$ ratio: particles with $V_{\text{rel}}/V_{\text{frag}}<1$ represent grains that increased in size during the last timestep, while particles with $V_{\text{rel}}/V_{\text{frag}}>1$ represent grains that fragmented. Initially, small particles tend to grow. As they are tightly coupled to the gas they have $|\Delta \bm{v}| \ll V_{\mathrm{t}}$ and $\mathrm{St}\ll1$, thus having low $V_{\text{rel}}$ \citep[see Figure 1 of][]{Vericel+2021}. As grains grow, they decouple from the gas and increase in $V_{\text{rel}}$ \citep{StepinskiValageas1997}. This goes on until $V_{\text{rel}}$ becomes close to $V_{\text{frag}}$ and the grains stop growing. The size of grains is then fragmentation-limited. Particles at the top of the grain size distribution having $V_{\text{rel}}/V_{\text{frag}}\lesssim1$ are in this fragmentation-limited regime. As seen in \textit{OB0}, \textit{OB5}, and \textit{ref}, the fragmentation limit corresponds to different grain sizes depending on the local conditions. Regions with higher dust densities allow grains to grow larger. Conversely, particles in lower-density environments, such as layers above the midplane or regions slightly interior to the disc edge, reach a smaller fragmentation-limited size than particles in the densest regions. This explains the large span in grain size that we observe.
In the \textit{IB} simulations, the dominant effect is the mixing induced by the binary. The dust density varies with time, and individual particles never reach a steady state. Instead, they grow and fragment in response to the binary perturbations. Because the maximum dust density remains stable, the maximum grain size also remains roughly constant.

\begin{figure}[ht]
\centering
\begin{center}
    \includegraphics[width=\columnwidth, trim={0cm 0cm 0cm 0cm},clip]{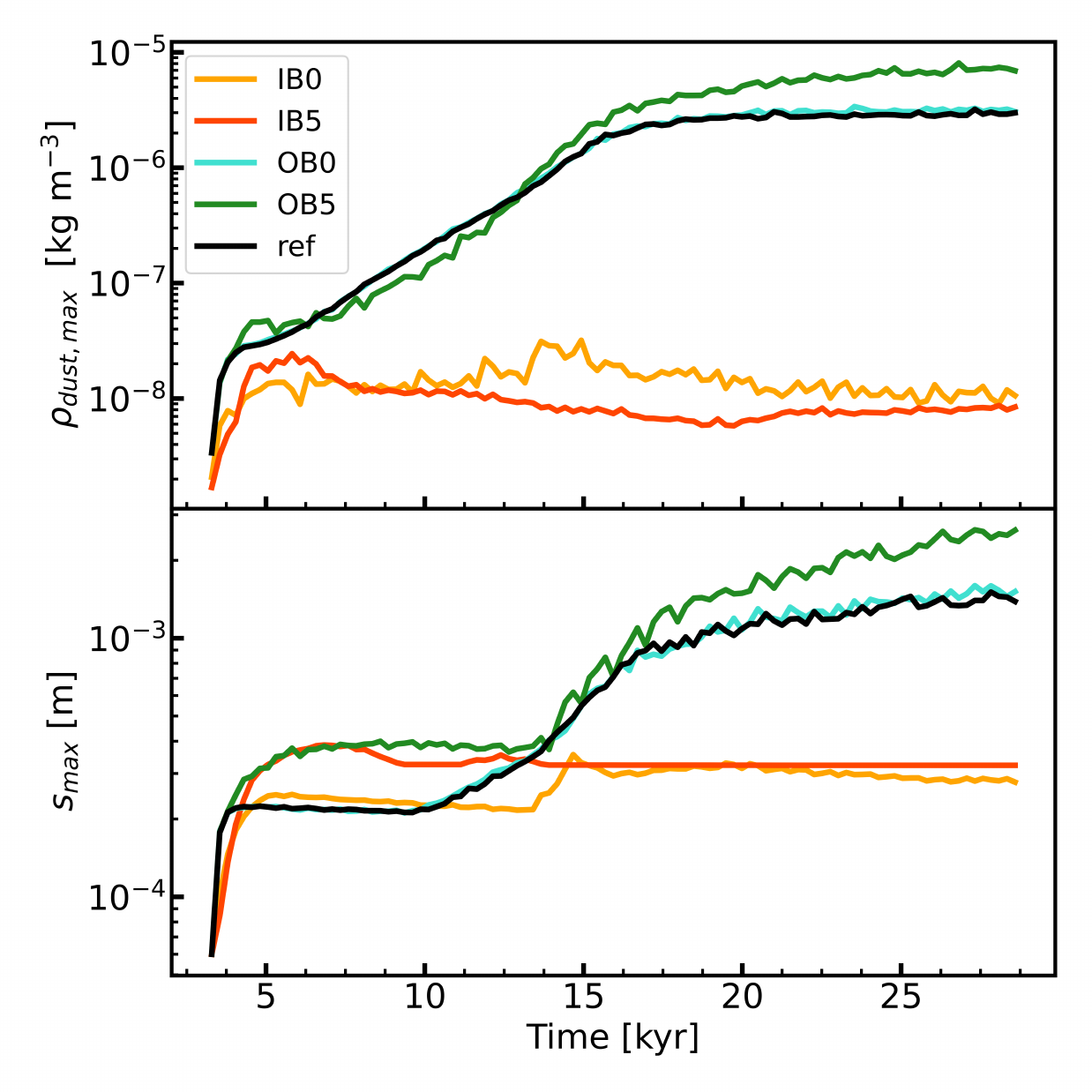}     
    \caption{
    Maximum dust density (top) and maximum grain size (bottom) in the disc as a function of time in each simulation. The simulation name is indicated at the top left of the top panel.}
    \label{fig:maxgrainsizemaxdust}
\end{center}
\end{figure}

Figure \ref{fig:maxgrainsizemaxdust} shows the maximum dust density and maximum grain size in the disc as a function of time. The maximum dust density quickly rises by one order of magnitude compared to the initial value. Then, it stagnates until the end of the \textit{IB} simulations while it rises continuously in the other simulations. We attribute this continuous rise to the replenishment of the inner disc by radial drift (see Appendix \ref{app : acc drift}). Towards the end of the simulations ($>20$~kyr), almost all the dust has reached the inner disc and the maximum dust density saturates. In circumbinary discs, such phenomenon is mitigated by the perturbations of the inner binary. Note that the maximum dust density is larger by a factor $\sim2.3$ in \textit{OB5} compared to \textit{ref}.
The maximum grain size follows the same trend \mdy{than}{as} maximum dust density. There is a sharp increase in grain size at the beginning of the simulations, which then flattens in the \textit{IB} cases, and continuously rises in the other cases. The maximum dust density and the maximum grain size are correlated, which follows from the size evolution rate being proportional to the dust density \citep{Vericel+2021}. 
These results show that, in our simulations, grain growth is mainly controlled by local dust density.

\subsection{Streaming instability}
\label{subsec:SI}

By evolving the dust grain size in our simulations, we tested the collisional growth of dust grains into pebbles in multiple-star systems. We now go a step further and assess the formation of planetesimals. 
To overcome the radial drift barrier \citep{Weidenschilling1977,TakeuchiLin2002}, planetesimal formation is thought to occur via rapid processes involving disc instabilities \citep{Lesur+2023}. One of these processes is the streaming instability, which consists \mdy{in an unstable feedback loop}{of a state of unstable feedback} between gas and dust \citep{YoudinGoodman2005}. We consider here the non-resonant streaming instability. Our goal is not to model the detailed development of this instability in binary systems, but rather to determine whether its triggering conditions are met in our simulations.
A key parameter in this respect is the dust-to-gas ratio, \mdy{noted}{denoted} $\delta_{\text{dg}}$. The non-resonant streaming instability develops when $\delta_{\mathrm{dg}}\gtrsim1$, at which point dust feedback becomes strong enough to perturb the gas disc \citep{SquireHopkins2018}.

\begin{figure*}[ht]
\centering
\begin{center}
    \includegraphics[width=\textwidth, trim={0cm 0cm 0cm 0cm},clip]{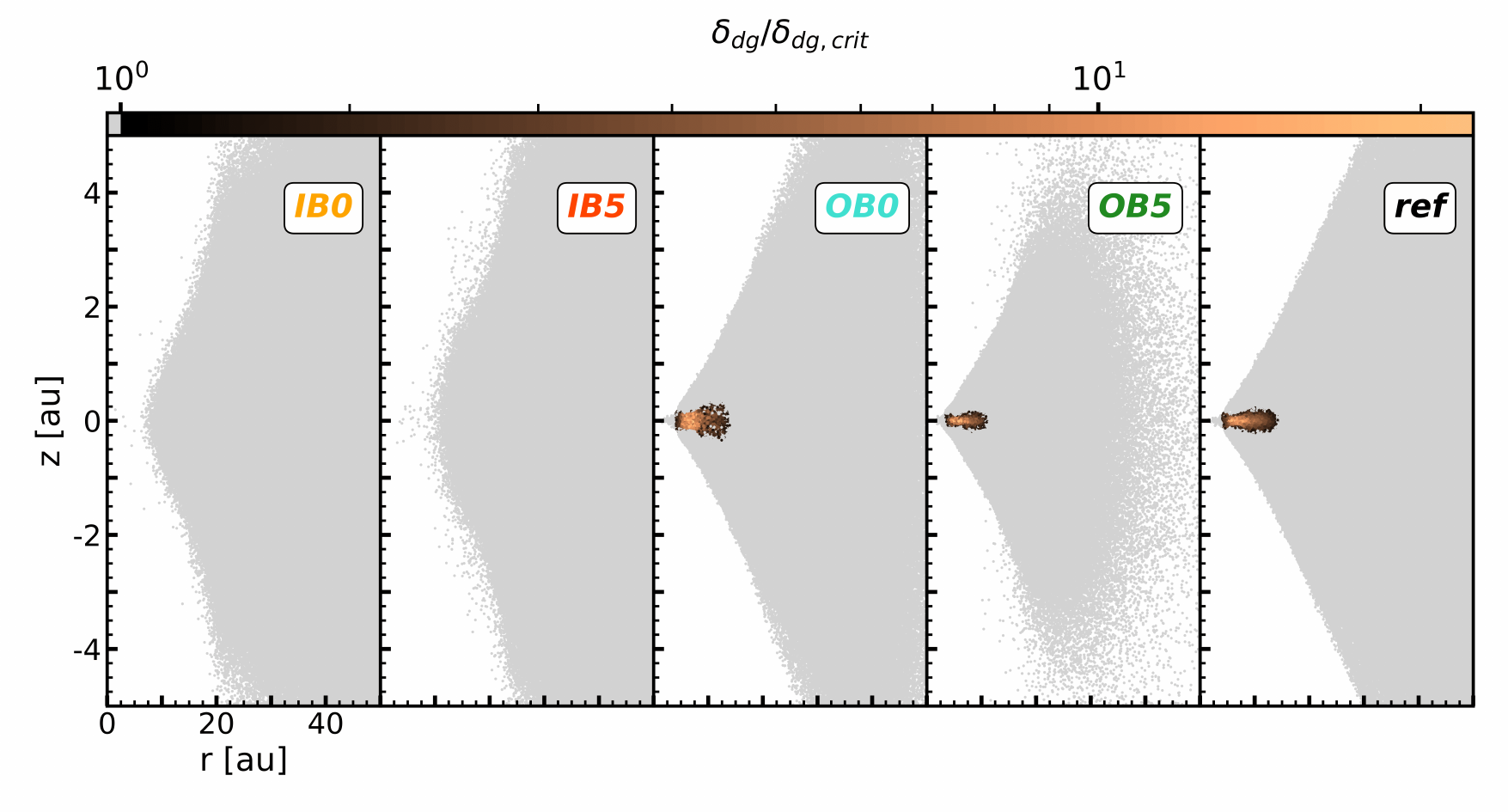}     
    \caption{Dust-to-gas ratio as a function of space in each simulation. Regions highlighted by coloured markers present favourable conditions to the development of the streaming instability and strong particle clumping, as given by Equation \ref{eq:eps_crit}. Conversely, strong clumping is not expected to develop in regions indicated by grey markers. The simulation name is indicated at the top right of each panel. \mdy{}{The panels display the full 3D data projected onto the plotting plane.}
    }
    \label{fig:SI_epscrit}
\end{center}
\end{figure*}

\cite{Lim+2024} gave a relation for the critical dust-to-gas ratio needed to trigger the streaming instability and sustain particle clumping, that we note $\delta_{\text{dg,crit}}$, as a function of the Stokes number. That relation is
\begin{equation}
 \label{eq:eps_crit}
\log \delta_{\text{dg,crit}} \approx 
0.42\log(\mathrm{St})^2+0.72\log(\mathrm{St})+0.37 \,.
\end{equation}
Figure \ref{fig:SI_epscrit} shows the spatial distribution of $\delta_{\text{dg}}$ for each simulation. Particles with $\delta_{\text{dg}}>\delta_{\text{dg,crit}}$ are coloured as a function of $\delta_{\text{dg}}$, while particles with $\delta_{\text{dg}}<\delta_{\text{dg,crit}}$ are plotted in grey. 
For circumbinary discs, we have $\delta_{\text{dg}}<\delta_{\text{dg,crit}}$ in the entire disc. These results suggest that, under our assumptions, circumbinary discs do not provide favourable conditions for triggering strong clumping by the streaming instability. This outcome can be interpreted in light of the vertical mixing shown in Figure \ref{fig:rz_dustdensity} (see also Appendix \ref{app : inner cbd}). Circumbinary discs exhibit larger gas scale heights and lower dust densities than circumstellar discs. As a result, the dust-to-gas ratio $\delta_{\text{dg}}$ remains lower than the onset threshold of strong clumping \citep{LiYoudin2021}. In addition, external turbulence, such as that driven by the binary in the inner disc, can further inhibit the clumping produced by the streaming instability \citep{Lim+2024}.

For circumstellar discs, we have $\delta_{\text{dg}}>\delta_{\text{dg,crit}}$ in a region close to the midplane ($z<1$~au) from the inner disc edge to approximately $15$~au. The regions favourable to streaming instability and strong clumping are of similar extent between \textit{OB0} and \textit{ref}, but are sensibly narrower in \textit{OB5}. The values of $\delta_{\text{dg}}$ peak at $\sim32\delta_{\text{dg,crit}}$ in all circumstellar cases.
If we compute the dust mass enclosed in the region favourable to the streaming instability and strong clumping, we obtain $40\pm6$~M$_{\oplus}$, $43\pm7$~M$_{\oplus}$, and $38\pm6$~M$_{\oplus}$ for \textit{OB0}, \textit{OB5}, and \textit{ref}, respectively.
These results indicate that, if dust density is the primary driver of dust growth, the sub-structures induced by external companions do not significantly alter the development of the streaming instability compared to isolated discs.

\section{Discussion}
\label{sec:discussion}

\subsection{Planet formation in circumbinary discs}
\label{subsec:disc_CBD}

Our simulations show that dust growth in circumbinary discs coplanar with their host binary seems limited compared to the same process in circumstellar discs. Considering our numerical setup, dust growth is mainly driven by the local dust density. In our circumbinary discs, dust density at the inner disc edge is almost four orders of magnitudes lower than in circumstellar discs. This is a consequence of the radial and vertical mixing induced by the inner binary, as extensively studied by \cite{Pierens+2021}. This mixing could originate from the turbulence following a parametric instability \citep{Papaloizou2005a,BarkerOgilvie2014} or from vortices formed at the inner disc edge \citep{CimermanRafikov2024}. In any case, this phenomenon complicates and limits the efficiency of dust grain growth: a result found by \cite{Pierens+2021} based on measurements of $V_{\text{rel}}$ from disc kinematics in grid simulations\footnote{In a follow-up paper, we will provide a method to measure $V_{\text{rel}}$ from SPH simulations.} that we corroborate in this work. An inefficient grain growth has consequences on the formation of planetesimals, as grain growth and dust accumulation help to trigger the streaming instability \citep[e.g.][]{Vallucci-Goy+2026}. \cite{Pierens+2021} suggested that circumbinary discs hinder the triggering of the streaming instability. While we followed an independent approach based on SPH simulations, our results are in agreement with their findings. We provide new elements of evidence that the interplay between radial drift, vertical settling, and dust growth is not sufficient against the perturbations \mdy{of}{caused by} the inner binary to create favourable conditions for strong clumping driven by the streaming instability close to the binary ($\lesssim5$~a$_{\text{bin}}$ for $a_{\text{bin}}\gtrsim5$~au). This statement is in line with the work of \cite{SilsbeeRafikov2015}, who also found that planetesimal growth was inhibited in the inner regions of circumbinary discs. It appears that planet formation, from the collisional growth of dust grains to the assembly of planetary cores, is unlikely to occur close to the cavity edge of circumbinary discs.

Yet, in evolved systems, numerous circumbinary planets (P-type, \citealt{Dvorak1982}) were detected close to the stability limit imposed by the inner binary ($\lesssim6$ P$_{\text{bin}}$, corresponding to $\lesssim3.3$ a$_{\text{bin}}$, \citealt{Baycroft+2024}). For example, out of the ten systems with at least one P-type planet detected by \textit{Kepler} \citep{Borucki+2010} until 2024, nine of them have a planet close to the stability limit \citep{Welsh+2014,Baycroft+2024}. While it remains debated whether this pile-up is an artefact or a genuine feature \citep{Armstrong+2014,MartinTriaud2014,Li+2016,Quarles+2018}, it questions the formation pathways of P-type planets detected near the stability limit: are these planets actually formed \textit{in situ} close to the binary, or do they form at larger separations and subsequently migrate inwards?
All the reasons cited earlier in this Section argue against an \textit{in situ} formation. An additional argument comes from the disc eccentricity at the location of the observed \textit{Kepler} P-type planets. The eccentricity of the disc is greater than that of the observed planets: our circumbinary discs have an eccentricity of $\gtrsim0.1$, while \textit{Kepler} P-type planets have an eccentricity of $\sim0.05$ (see Appendix \ref{app : disc size} and Appendix \ref{app : cbp}). This mismatch implies that, if planets were to form \textit{in situ}, they would need to undergo substantial orbital circularisation after disc dispersal, for which no efficient mechanism is readily apparent \citep{PelupessyPortegies2013,Zoppetti+2020}. As a consequence, it seems unlikely that planets close to the stability limit have formed \textit{in situ} following the core accretion scenario. 

Instead, P-type planets may rather form far away from the inner binary, in less perturbed regions, and then migrate inward \citep[e.g.][]{ThunKley2018,TeasdaleStamatellos2023}. In the core accretion scenario, the larger the grains grow, the faster their radial drift. Without sub-structures to halt this drift, the outer regions of protoplanetary discs are depleted of dusty material ready to form planetesimals. Consequently, forming planets in these regions far from the inner binary is unlikely within the core accretion scenario. This conclusion does not exclude planet formation at large distances from the binary in general. In our \textit{IB} simulations, the dust density starts to decrease at separations of $\sim6$~a$_{\text{bin}}$. At separation of $6$~a$_{\text{bin}}$ around a more compact binary, the dust density would however be higher than in our case, possibly providing favourable conditions for planet formation. Such conditions could allow the formation of the transiting \textit{Kepler} circumbinary planets near their observed separation, which is in average $4.97\pm0.02$~a$_{\text{bin}}$ (see Appendix \ref{app : cbp}). Exploring this regime is left for future work.

To form planets in the outer disc, a possible pathway relies on the gravitational instability \citep{Boss1997,Gammie2001}. In that context, circumbinary discs may present more favourable conditions to the gravitational instability compared to isolated discs \citep{TeasdaleStamatellos2026}. Gravitational instability provides a natural explanation for the giant P-type planets detected at large separations \citep{ThebaultBonanni2025}. Indeed, the mean mass of circumbinary planets detected by direct imaging is $10\pm1$~M$_{\text{J}}$ for an average separation of $118\pm8$~a$_{\text{bin}}$ (see Appendix \ref{app : cbp}). 
Explaining the transiting \textit{Kepler} circumbinary planets remains more challenging. These planets are much less massive, with an average mass of $0.06\pm0.47$~M$_{\text{J}}$, and orbit much closer to the binary. Such properties are difficult to reconcile with formation by gravitational instability, which typically produces planets of $1-100$~M$_{\text{J}}$ at several tens of au \citep{Boley+2011,Adams+2025}.

\subsection{Planet formation in circumstellar discs in binary systems}
\label{subsec:disc_CSD}

Our simulations show comparable levels of grain growth between circumstellar discs in isolated or in binary systems. In our framework, this can be explained by a similar evolution of the dust density between the two setups. This evolution is mainly driven by the radial drift of growing grains, which accumulate at gas pressure bumps, namely at the inner disc edge in our simulations. In case of an eccentric binary, radial drift is enhanced \citep{Manara+2019,Rota+2022}, which helps the pile-up of dust. Once all the grains have reached the pressure bump, we would witness the formation of a dense dust ring in which the size of dust grains is limited by fragmentation \citep{Gonzalez+2017,Vericel+2021}. From an observational perspective, however, it remains challenging to disentangle the formation pathways of such rings from other mechanisms capable of trapping dust \citep{Bae+2023}.

Close to the inner disc edge ($<15$~au) and in the midplane, we find favourable conditions for the streaming instability to lead to strong particle clumping. We find no difference within the error-bars in the mass convertible into planetesimals by the streaming instability when comparing isolated discs to circumstellar discs in binaries. However the evolution and growth of the planetesimals may differ between those cases. Indeed, the gravitational pull of the companion increases the impact velocities of planetesimals \citep{Paardekooper+2008}. As a consequence, collisions predominantly result in erosion at separations that can reach the inner regions ($\gtrsim0.05$~a$_{\text{bin}}$ for $a_{\text{bin}}=23.4$~au,  \citealt{Thebault+2009}), preventing the steady growth of planetesimals by coagulation. This points out that planetesimals in externally perturbed discs may instead be born large ($\gtrsim10$~km) from rapid processes: the streaming instability is able to form such large planetesimals \citep{Schafer+2017}, which are then able to survive and grow in circumstellar discs in binaries \citep{SilsbeeRafikov2021}.

The inner disc regions are the most prone to form planets via core accretion. Population studies indeed found that circumstellar planets in binaries (S-type) preferentially form close to their host star ($a<0.1a_{\text{bin}}$, \citealt{Venturini+2026,Nigioni+2026}). In addition, if we note $a_{\mathrm{p}}$ the semi-major axis of the planet and $a_{\text{crit}}$ the semi-major axis over which S-type planets are expected to become unstable, we find $a_{\mathrm{p}}/a_{\text{crit}}=0.026\pm0.134$ over the population of detected S-type planets\footnote{from the catalogue of \href{https://exoplanet.eu/planets_binary/}{exoplanet.eu}, assuming coplanar binaries and the stability limit from \cite{HolmanWiegert1999}. In case of a multi-planet system, we only considered the innermost planet.}. Even if observational biases may mitigate this conclusion, the location of S-type planets coincides with the area in the disc favourable to their formation.

External stellar companions tidally truncate protoplanetary discs \citep[e.g.][]{ArtymowiczLubow1994}. Tidal truncation limits the radial extent of the disc and therefore reduces the area available for planet formation. In addition, this process may eject material from the disc or trigger accretion onto the central star, thereby decreasing the mass reservoir from which planets can form \citep{Cuello+2019}.
We find that the dust mass convertible into planetesimals via the streaming instability (see Section \ref{subsec:SI}) is of comparable order in truncated circumstellar discs in binaries and in discs around single stars. When tidal truncation does not affect the regions susceptible to the streaming instability, it therefore plays only a minor role in setting the dust reservoir available for planet formation. Our results also show that, for discs of similar size, density-driven dust growth proceeds similarly regardless of the presence of an external companion. In that case, planet formation would be likely driven by the disc’s initial dust mass \citep{Mishra2023}.

External stellar companions drive large-scale spiral arms, locally compressing and accelerating the gas and creating pressure maxima in the disc. Dust grains that are well-coupled to the gas ($\mathrm{St}<1$) can be efficiently entrained within the spirals and concentrated by the converging gas flows. This enhancement in local dust density may promote the coagulation of dust particles. While dust particles are entrained by the spirals in our simulations, this does not result in a significant accumulation of dust (see Figure \ref{fig:all_density}). Over time, the bulk of the dust particles drifts towards the inner regions, progressively reducing the magnitude of the spirals' feedback. As a consequence and as observed in our simulations, spiral density waves in circumstellar discs do not promote the collisional growth of dust grains. It would be interesting to see if decoupled grains ($\mathrm{St}>1$) can be concentrated in circumstellar discs in binaries as in flybys \citep{Prasad+2025,Su+2026} and further grow.

\subsection{Caveats}
\label{subsec:discussion}

Some of our results may depend on the specific orbital configurations adopted in the simulations. We explore here to what extent they can be scaled with the binary semi-major axis.
For circumbinary discs, the regions located at a few $a_{\text{bin}}$ from the binary are perturbed and turbulent, and therefore unfavourable to planet formation. In theory, this does not prevent grain growth and streaming instability to occur further out in the disc. As $a_{\text{bin}}$ decreases, the regions close to the cavity becomes denser. It is not clear to what extent this would allow dust growth to proceed, and if there is a smooth transition between \textit{IB} and \textit{ref} as $a_{\text{bin}}$ decreases. Conversely, these considerations imply that dust growth is inhibited if $a_{\text{bin}}>5$~au.
For circumstellar discs in binaries, the companion has a low mass and follows a relatively wide orbit. This choice guarantees the presence of structure in the disc and a viable comparison with the single-star case. A closer, more eccentric, or more massive companion in the \textit{OB} simulations would truncate the disc more strongly, and the comparison with the \textit{ref} simulation would no longer hold. Under the assumption that dust growth is driven by local density, our results are then scalable as long as the inner disc regions are marginally perturbed by the companion.
Determining the separations at which external companions inhibit density-driven dust growth, or inner binaries allow it, requires a systematic exploration of the orbital parameter space. We leave this investigation to future work.

The time over which the simulations are integrated may not be long enough to reach a steady state. As shown by Figure \ref{fig:maxgrainsizemaxdust}, the maximum dust density and grain size saturate in \textit{IB0} and \textit{IB5}, which we associate to a steady state. In \textit{OB0}, \textit{OB5}, and \textit{ref}, the maximum grain size is still steadily increasing at the end of the simulation because of the replenishment of the inner disc by radial drift. In such case, the impact of the companion on dust growth is marginal, and dust growth is rather governed by the disc physics. As a consequence, the simulations of circumstellar discs are in a steady state with respect to the perturbations of the companion, the global steady state being reached once all the grains have drifted.

Different values of $V_{\text{frag}}$ would have changed the outcome of grain growth and fragmentation in our simulations. However, the results obtained by comparison between the different simulations would still hold. We stress that our choice of $V_{\text{frag}}=15$~m~s$^{-1}$, representing grains coated with icy materials, suits well to this study: in circumbinary discs, the water snowline lies close to the cavity edge and previous estimations of $V_{\text{frag}}$ match our choice \citep{PierensNelson2024,Alaguero+2025}. In circumstellar discs, the bulk of the disc is beyond the water snowline, the position of which is hardly impacted by external companions \citep{Poblete+2025}.

Another limitation is the assumption that microscopic gas turbulence is the main driver of the collisions between dust grains. The dust growth formalism used in this work does not take into account the feedback of the disc kinematics on the relative velocity of dust grains. Grain collisions due to non-Keplerian motions are believed to be important in multiple-star systems. For example in circumbinary discs, $V_{\text{rel}}$ is found larger than the relative velocities driven by microscopic gas turbulence at the cavity edge \citep{Pierens+2021}. To cope with that, we will propose an new way to calculate $V_{\text{rel}}$ in SPH in a companion paper (Alaguero et al. in prep).
In this work, we do not attempt at giving quantitative predictions of the dust grain size in multiple systems, but we instead try to capture the impact of the perturbed density structure on dust growth. 
The grain sizes obtained in this work should be regarded as upper limits. Indeed, the gravitational stirring produced by stellar companions would increase $V_{\text{rel}}$ compared to this work, which would result in lower grain sizes.


\section{Conclusion}
\label{sec:conclusion}

In this work, we studied the impact of the density structure of discs in binary systems on the growth of dust grains. To this end, we performed a suite of hydrodynamical simulations of discs orbiting one or both stellar components of a binary system, explicitly modelling dust growth and fragmentation processes driven by gas turbulence. Our main findings are summarised as follows:
\begin{itemize}
    \item The density structure of discs is significantly shaped by tidal interactions with the companion star. In circumbinary discs, dust is trapped at the cavity edge, where it is stirred radially and vertically. In circumstellar discs within binary systems, dust rapidly drifts inwards, settles towards the midplane, and accumulates in the inner regions of the disc. These processes are further enhanced when the companion follows an eccentric orbit.
    \item From that density structure, the maximum grain sizes reached in circumbinary discs are $\sim5$ times lower than in isolated discs. This originates from the perturbations of the binary on the inner disc. In circumstellar discs in binaries following a circular orbit, maximum grain sizes are comparable to isolated discs. In case of an eccentric external companion, the maximum dust density is twice higher, which is converted into grain sizes approximately twice larger.
    \item Conditions favourable to the onset of the streaming instability and strong dust clumping are met in circumstellar discs with external companions, similarly to isolated discs. The regions favourable to the instability match the observed location of S-type planets. In circumbinary discs, however, the conditions for the streaming instability to produce strong clumping are not satisfied at any location. It suggests that P-type planets observed close to the dynamical stability limit may have formed at larger separations and further migrated to their current location.
\end{itemize}

We emphasise that these results were obtained while neglecting the impact of large-scale, non-Keplerian disc kinematics on dust grain growth. This limitation will be addressed in a forthcoming paper, in which we will incorporate the feedback of disc kinematics on the relative velocities of dust grains. This follow-up study will allow us to disentangle the respective roles of dust density and disc dynamics in regulating grain growth, and to provide quantitative predictions for dust grain sizes in protoplanetary discs within binary systems.


\begin{acknowledgements}
    We would like to express our gratitude to the anonymous referee for their comments and suggestions which have greatly contributed to the clarity of this work.
    AA express his sincere gratitude to Fran\c cois Ménard for his insightful comments. This project has received funding from the European Research Council (ERC) under the European Union Horizon Europe programme (grant agreement No. 101042275, project Stellar-MADE). AA would like to thank the members of the Stellar-MADE team that are not in the author list of this work (Mario Sucerquia, Pedro Poblete, Elisa Castro, and Romain Grane) for their scientific support during this project.
    JFG thanks the LABEX (LABoratoire d’EXcellence) Lyon Institute of Origins (ANR-10-LABX-0066) for its financial support within the Plan France 2030 of the French government operated by the ANR (Agence Nationale de la Recherche).
    ML acknowledges funding from European Research Council (ERC) under the European Union’s Horizon 2020 research and innovation program (Grant agreement No. 101098309 - PEBBLES).
    This work makes use of {\sc splash} \citep{splash}, {\sc Sarracen} \citep{Sarracen}, {\sc Numpy} \citep{numpy}, and {\sc Matplotlib} \citep{Matplotlib}. 
\end{acknowledgements}

\bibliographystyle{aa} 
\bibliography{biblio}

\begin{appendix}
\nolinenumbers

\section{Stokes number and sanity checks}
\label{app : St}

\begin{figure*}[b]
\centering
\begin{center}
    \includegraphics[width=\textwidth, trim={0cm 0cm 0cm 0cm},clip]{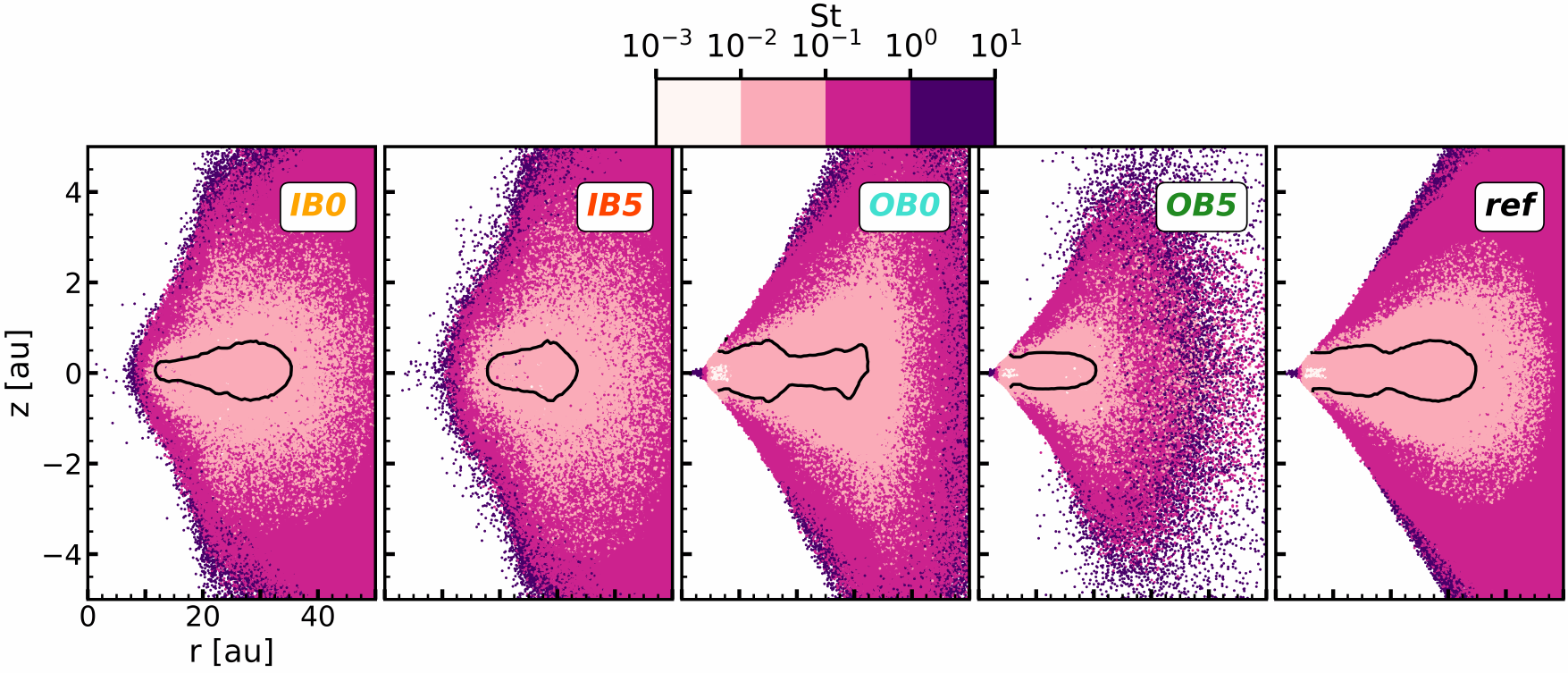}     
    \caption{Stokes number as a function of space for each simulation. The black contour delineates the $\delta_{\text{dg}}>0.01$ region. The simulation name is indicated at the top right of each panel. \mdy{}{The panels display the full 3D data projected onto the plotting plane.}
    }
    \label{fig:St}
\end{center}
\end{figure*}

The hydrodynamical simulations of this work include gas and dust interacting together. The coupling between gas and dust was treated using a dust-as-a-mixture approach \citep{LaibePrice2014c,LaibePrice2014a,LaibePrice2014b,PriceLaibe2015,Ballabio+2018}. This formulation breaks down when the dust velocity field becomes multi-valued, in case of particles with intersecting trajectories for instance. \mdy{Then}{Therefore}, as long as the dust follows the gas flow, the dust-as-a-mixture approach remains valid. In terms of dust-gas coupling, this is equivalent to verify $\mathrm{St}<1$.
Figure \ref{fig:St} shows the Stokes number of each particle as a function of space for each simulation. On that figure, the black contour delineates the $\delta_{\text{dg}}>0.01$ region. We see that the bulk of the particles have $\mathrm{St}<1$, verifying the working condition for the dust-as-a-mixture scheme. Some particles have $\mathrm{St}>1$, but they are located close to the disc surface and outside of the $\delta_{\text{dg}}>0.01$ region. Those areas are depleted of dust and are of limited interest when considering dust growth.

\section{Disc size}
\label{app : disc size}

In this \mdy{paragraph}{appendix}, we provide details about the disc extent. In binary systems, tidal truncation significantly impacts the disc size. The gas disc size is then set by the outcome of the competition between tidal truncation and viscous spreading. The dust disc size depends on tidal truncation and on the coupling with the gas, which affects how dust spreads and drifts. In general, the disc size is greater in the gas than in the dust because of radial drift \citep{Weidenschilling1977}.
In this Section, we measured the disc outer radius $R_{\mathrm{out}}$ as the radius encircling $90\%$ of the particles in an attempt to match observational methods. Given the non-zero eccentricity of the cavity in the \textit{IB} simulations, we used the cavity semi-major axis $a_{\text{cav}}$, defined as the semi-major axis where density reaches half of its peak value, to measure the location of the inner disc edge. The error on $a_{\text{cav}}$ was taken as the difference between the half-peak value and the quarter-peak value. The error on $R_{\mathrm{out}}$ was taken as the azimuthally averaged smoothing length in the corresponding radial bin. Measurements of the gas extent were done using all the particles in the simulations, while only particles with $\delta_{\mathrm{dg}}>0.01$ were considered to calculate the extent of the dust disc. Initially, all the particles have $\delta_{\mathrm{dg}}=0.01$, but dust quickly drifts towards the inner half of the disc, making outer regions having $\delta_{\mathrm{dg}}<0.01$.

At $t=0$ for the \textit{IB} simulations, we measured $a_{\text{cav}}=9.1\pm3.3$~au. For all the simulations, we measured $R_{\mathrm{out}}=38.0\pm0.2$~au for the initial condition. Note that dust and gas are initially equally distributed. Table \ref{table:radii} summarises the extent of the simulated discs at $t=29$~kyr. 
At the inner rim, the central binary has carved a cavity with time. The cavity is found smaller in gas than in dust, likely because the dust particles get stopped at the gas pressure bump while the gas can flow in the cavity (see Appendix \ref{app : acc drift}). The cavity is larger in \textit{IB5} than in \textit{IB0}, which is theoretically expected. Indeed, tidal truncation theory indicates truncation radii of approximately $7.5$~au for \textit{IB0} and of approximately $14.5$~au for \textit{IB5} \citep{MirandaLai2015}. These predictions are in the range delimited by the inner edge of the dust and gas discs. 
Eccentric cavities around binaries are expected to develop \citep{Hirsh+2020,Ragusa+2020}. The gas cavities in \textit{IB0} and \textit{IB5} are respectively found with $e=0.14$ and $e=0.13$, approximately.

Let us now investigate circumstellar discs in binary systems. The dust disc outer radius is similar in \textit{OB0} and \textit{ref}, showing that the size of the dust disc in \textit{OB0} is set by radial drift rather than tidal truncation. However, the external companion limits the extent of the gas disc. In \textit{OB5}, $R_{\mathrm{out}}$ is naturally found lower than in the other simulations.
We estimated the truncation radius using the empirical formula of \cite{Manara+2019} for $b=-0.75$ and $c=0.68$\footnote{\textit{i.e.} $\alpha_{\mathrm{SS}}=5\times10^{-3}$, $H/R=0.05$, and $\mu=M_2/M_1=0.1$.}, which gives a truncation radius of $68$~au for \textit{OB0} and of $31$~au for \textit{OB5}. 
We obtained more accurate estimates by using the tidal truncation theory, which indicates outer truncation radii of approximately $40$~au for \textit{OB0} and of approximately $22$~au for \textit{OB5} \citep{MirandaLai2015}. These predicted values are in good agreement with the extent of the gas disc of the \textit{OB} simulations. Comparing the dust-to-gas disc size ratio with the literature can also provide insights on the dust-gas dynamics. In Taurus, circumstellar discs around primary stars are found with dust-to-gas size ratios between $0.1$ and $0.5$ \citep{Manara+2019,Rota+2022}. This ratio is of $\sim0.74$ in the simulations, which does not match the observational data. Letting the simulations evolve for longer times would allow for the dust to drift to smaller radii and possibly reproduce observational data. If the dust grains were to grow in the process, they would get closer to $\mathrm{St}=1$ and it would accelerate their radial drift.

\begin{table}
    \centering
    \caption{Inner and outer edges of the simulated discs at the end of the simulations.}
\begin{tabular}{c| c c| c c}
             & \multicolumn{2}{c}{Dust} & \multicolumn{2}{c}{Gas} \\
             & $a_{\text{cav}}$ (au) & $R_{\mathrm{out}}$ (au) & $a_{\text{cav}}$ (au) & $R_{\mathrm{out}}$ (au) \\
 \hline\hline
\textit{IB0} & $9.1\pm0.6$ & $31.4\pm0.3$ & $7.5\pm1.3$ & $55.3\pm1.1$ \\
\textit{IB5} & $15.2\pm0.4$ & $31.5\pm0.2$ & $11.7\pm1.7$ & $55.1\pm0.7$ \\
\textit{OB0} & $5.3\pm0.2$ & $28.4\pm0.3$ & $4.5\pm0.3$ & $37.5\pm0.3$ \\
\textit{OB5} & $3.9\pm0.2$ & $16.6\pm0.2$ & $3.2\pm0.2$ & $22.7\pm0.3$ \\
\textit{ref} & $5.4\pm0.2$ & $28.7\pm0.3$ & $4.6\pm0.3$ & $52.3\pm0.7$
\end{tabular}
\tablefoot{$a_{\text{cav}}$ is the semi-major axis of the central cavity, defined as the semi-major axis where the density reaches half its peak value. $R_{\mathrm{out}}$ is the outer disc radius defined as the radius encircling $90\%$ of the particles. For $a_{\text{cav}}$, the errors are computed as the difference with $a_{\text{cav}}$ and the quarter-peak semi-major axis. For $R_{\mathrm{out}}$, the errors are taken as the azimuthally averaged smoothing length at $R_{\mathrm{out}}$.}
\label{table:radii}
\end{table}

\section{\mdy{Accumulation of drifting grains}{Dust accumulation in the outer disc}}
\label{app : acc drift}

Dust grains radially drift in protoplanetary discs as a consequence of gas drag \citep{Weidenschilling1977}. This drift is directed towards the location of pressure maxima \citep{Nakagawa+1986}. For this reason, pressure maxima are favourable locations for dust to accumulate. Grains grow over the course of their drift \citep{Laibe+2014-drift}. If dust grains grow fast enough to decouple from the gas ($\mathrm{St}\gg1)$ before reaching the disc inner edge, they may stop drifting and start accumulating in the disc: the dust back-reaction on the gas then leads to the formation of a pressure maximum, stopping the drift of incoming grains \citep{Gonzalez+2017}. 

In our simulations, however, grains remain in the $\mathrm{St}<1$ regime. They therefore do not decouple strongly from the gas and cannot form self-induced dust traps through this mechanism. An alternative pathway arises when grains drift at nearly constant size. As they migrate inward, they encounter progressively higher gas densities, which increases their aerodynamic coupling to the gas and lowers their Stokes number. Their drift then slows, making local dust accumulation more likely. We indeed see \mdy{on}{in} Figure \ref{fig:St} that $\mathrm{St}$ decreases when $r$ decreases, which is consistent with that scenario.
In the following, we study the drift of dust grains and to what extent they accumulate in our simulations.

\begin{figure*}[ht]
\centering
\begin{center}
    \includegraphics[width=\textwidth, trim={0cm 0cm 0cm 0cm},clip]{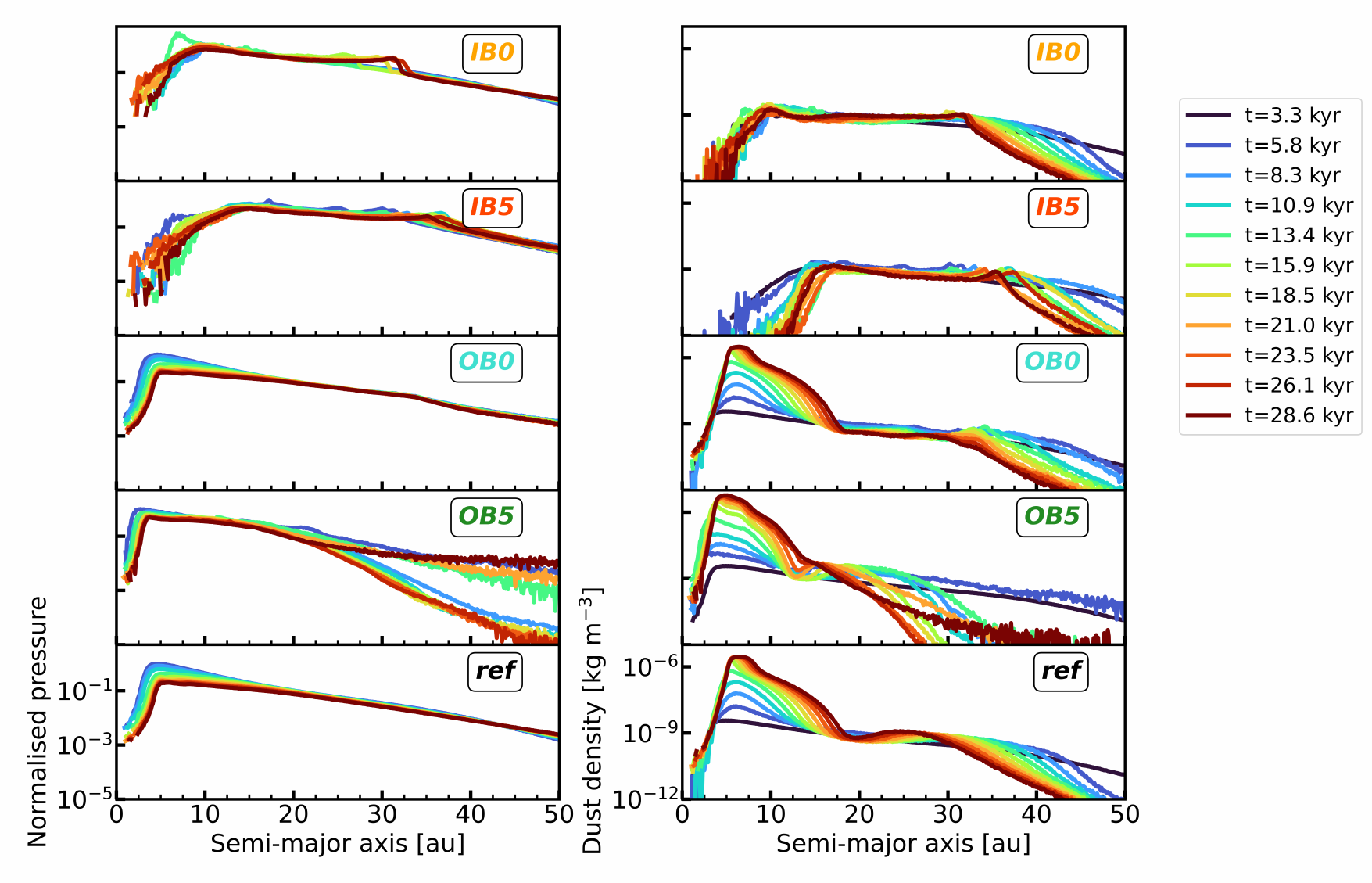}     
    \caption{\textbf{Left:} Radial profile of the normalised pressure as a function of time. The normalisation factor is arbitrarily taken as the pressure maximum at $5.8$~kyr. \textbf{Right:} Radial profile of the dust density as a function of time. The initial timestep, at $3.3$~kyr, corresponds to the moment the dust was added to the simulations. The radial profiles are computed by azimuthally averaging the data in concentric rings.
    }
    \label{fig:p_vs_radius_dustdens}
\end{center}
\end{figure*}

Figure \ref{fig:p_vs_radius_dustdens} shows the radial profiles of the gas pressure and of the dust density for each simulation as a function of time.
We observe radial drift in all the simulations. This can be seen by looking at the tail of the dust density distribution, which is moving inwards as a function of time\footnote{in simulations with eccentric companions, namely \textit{IB5} and \textit{OB5}, the location and slope of the tail also vary as a function of the orbital phase}. 
All the discs present a first peak in dust density at their inner radius. The presence of a second peak depends on the simulation.
In \textit{ref}, there is a second density peak at $\sim27$~au. From the temporal evolution, it is clear that this second peak is formed following the drift of dust grains. Yet, it has no counterpart in the pressure profile. As a consequence, the dust at that location may continue to drift inwards, or accumulate even more to decouple from the gas.
\mdy{}{In \textit{OB0}, a second density peak around $\sim32$~au was created throughout the course of the simulation, but that peak then flattened. This had formed a density plateau between approximately $20$~au and $30$~au. In \textit{ref}, this corresponds to the location where the dust density locally peaks. Then, this plateau is likely a result from tidal interactions with the companion: beyond $20$~au, drifting grains accumulate in \textit{ref} while they start to be cleared out by tidal truncation in \textit{OB0}. There seems to be a slight shift in the derivative of the pressure profile around $33$~au, but confirming this shift is at the limit of the data.
In \textit{OB5}, both radial drift and tidal truncation trim the dust spatial distribution. The profiles varies with time as a function of the phase of the companion: at periastron, the companion sends material in the external regions of the disc (see snapshot in Figure \ref{fig:all_density}), while at apoastron these structures dissipate and the disc density sharply decreases with at large $r$. Between $13$~kyr and $26$~kyr, there is a second peak in the dust density profile at $\sim17$~au, but there seems to be no pressure bump at that location. As for \textit{ref}, that dust peak may continue to drift inwards, or may gather enough dust to decouple from the gas and stop its radial drift.}

\begin{figure*}
\sidecaption
  \includegraphics[width=12cm]{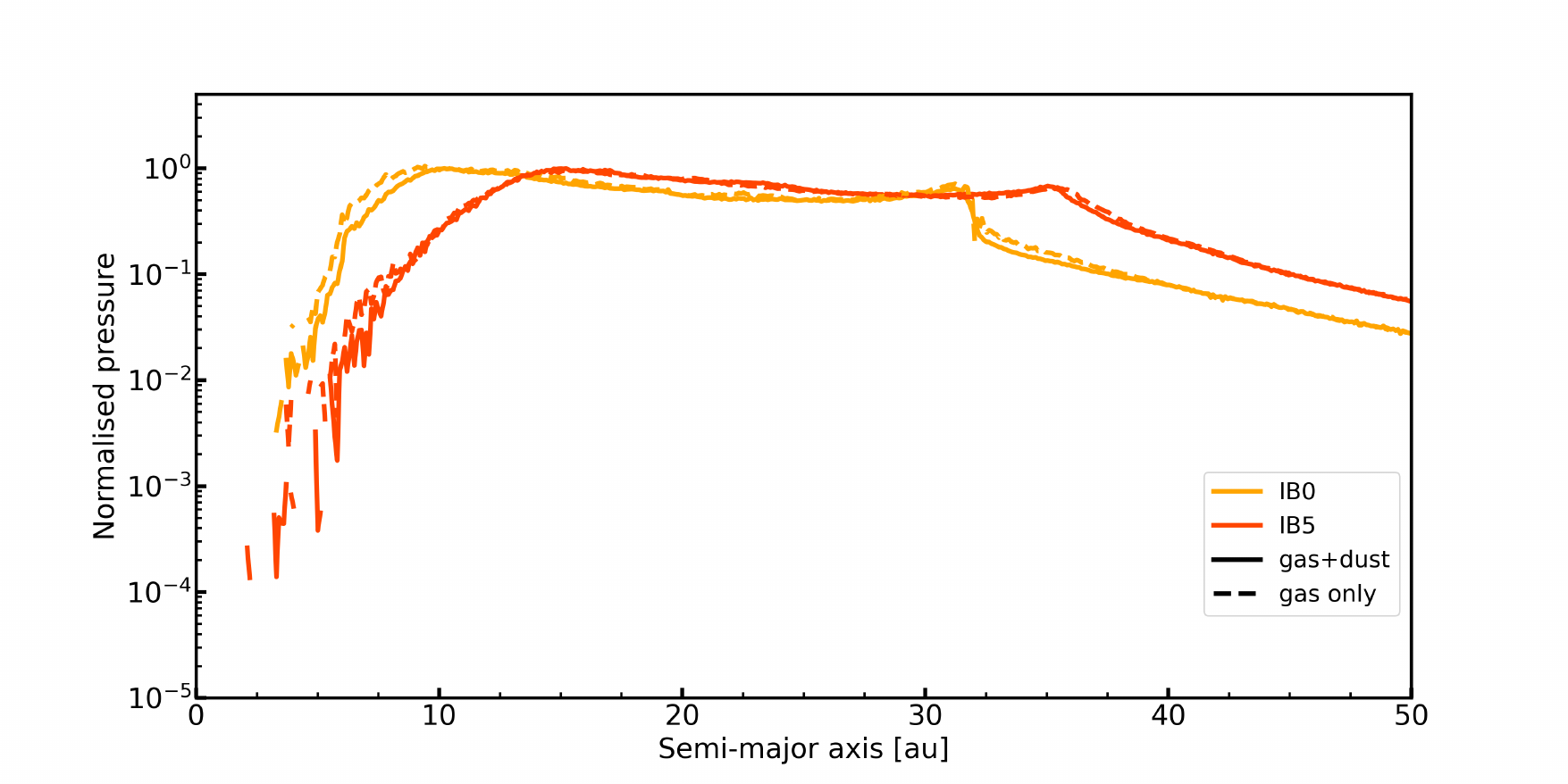}
     \caption{Radial profile of the normalised pressure as a function of time for \textit{IB0} and \textit{IB5} at $29$~kyr. The solid line corresponds to the simulations including growing dust, and the dashed line to simulations with gas only. The normalisation factor is arbitrarily taken as the maximum value in the gas+dust simulations. The radial profiles are computed by azimuthally averaging the data in concentric rings.}
     \label{fig:p_gasonly}
\end{figure*}

In \textit{IB0} and \textit{IB5}, there is a second density peak that has a counterpart in the pressure profile. These peaks are located at $\sim31$~au for \textit{IB0} and $\sim35$~au for \textit{IB5}. In circumbinary discs, we then deduce that the dust in the outer regions has drifted and is trapped at the location of the second density peak. \mdy{By telling the difference with \textit{ref}, this highlights the impact of the inner binary}{A direct comparison with \textit{ref} highlights the impact of the inner binary} on the drift of dust grains. \mdy{}{However, it remains unclear whether the pressure bump arises solely as a consequence of the inner binary or is instead initiated by dust back-reaction. To distinguish between these possibilities, we performed gas-only simulations of the \textit{IB0} and \textit{IB5} setups, keeping all other parameters identical to those of the original gas-dust simulations. Figure \ref{fig:p_gasonly} compares the pressure profiles obtained for the gas-only and gas-dust simulations. The two profiles are in excellent agreement. In particular, the pressure bump located near the outer edge of the disc is also present in the absence of dust. This demonstrates that the pressure bump is generated by the binary rather than the dust back-reaction. Based on these results, we propose the following scenario for the formation of the pressure bump. Viscous turbulence drives a net loss of angular momentum from the gas. At the same time, the inner binary excites spiral density waves that transport angular momentum outwards through the disc. As these waves propagate, they are progressively damped and deposit angular momentum into the surrounding gas. At the radius where the angular momentum supplied by the damped waves balances the angular momentum removed by viscous transport, the net radial motion of the gas is suppressed, allowing material to accumulate. This accumulation produces a local pressure maximum.
We emphasise that this interpretation remains speculative and that dedicated studies will be required to characterise the underlying mechanism and assess its role in the evolution of circumbinary discs.}

\mdy{In \textit{OB0}, a second density peak around $\sim32$~au was created throughout the course of the simulation, but that peak then flattened. This had formed a density plateau between approximately $20$~au and $30$~au. In \textit{ref}, this corresponds to the location where the dust density locally peaks. Then, this plateau is likely a result from tidal interactions with the companion: beyond $20$~au, drifting grains accumulate in \textit{ref} while they start to be cleared out by tidal truncation in \textit{OB0}. There seems to be a slight shift in the derivative of the pressure profile around $33$~au, but confirming this shift is at the limit of the data.
In \textit{OB5}, both radial drift and tidal truncation trim the dust spatial distribution. The profiles varies with time as a function of the phase of the companion: at periastron, the companion send material in the external regions of the disc (see snapshot in Figure \ref{fig:all_density}), while at apoastron these structures dissipate and the disc density sharply decreases with at large $r$. Between $13$~kyr and $26$~kyr, there is a second peak in the dust density profile at $\sim17$~au, but there seems to be no pressure bump at that location. As for \textit{ref}, that dust peak may continue to drift inwards, or may gather enough dust to decouple from the gas and stop its radial drift.}{}

\section{Turbulence in the inner regions of circumbinary discs}
\label{app : inner cbd}

The inner regions of circumbinary discs are known to be subject to an enhanced turbulence \citep{Pierens+2020}. In this Section, we qualitatively show that the circumbinary discs simulated for this study are also subject to an enhanced level of turbulence. 

\begin{figure*}[ht]
\centering
\begin{center}
    \includegraphics[width=\textwidth, trim={0cm 0cm 0cm 0cm},clip]{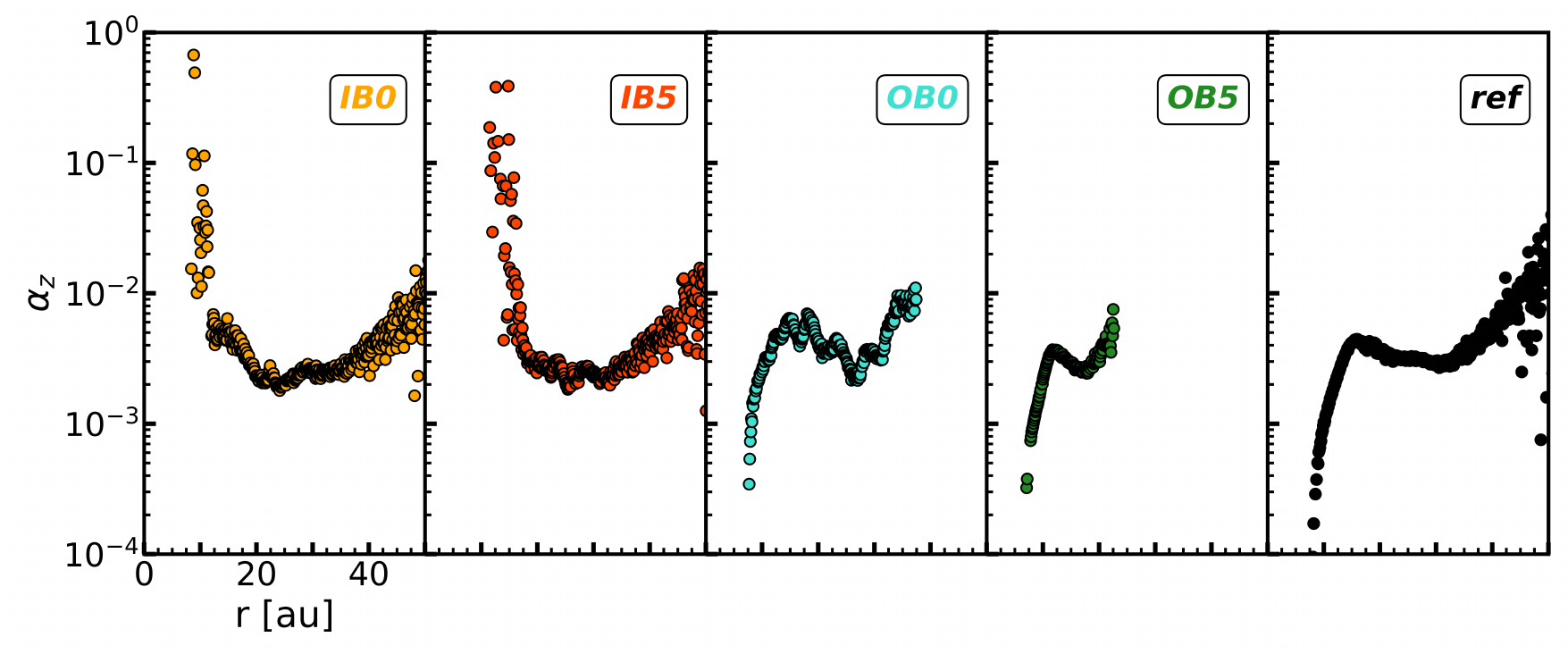}     
    \caption{Parametrisation of the vertical turbulence as a function of radius for each simulation at $29$~kyr. \mdy{}{The simulation name is indicated at the top right of each panel. These radial profiles were computed using Equation \ref{eq:alphaz} and azimuthally averaged profiles of $\text{St}$, $H_{\mathrm{d}}$, and $H_{\mathrm{g}}$. }
    }
    \label{fig:alphaz}
\end{center}
\end{figure*}

In the framework of \cite{ShakuraSunyaev1973}, we estimated the parametrisation of the vertical turbulence level, that we note $\alpha_{\mathrm{z}}$. \mdy{To do this}{To this end}, we first needed to assume balance between vertical stirring and settling. \mdy{This condition was reasonably checked by our simulations}{Upon verification, we checked that this condition was reasonably met in our simulations}, that evolved until equilibrium of the maximum dust density. In that context, $\alpha_{\mathrm{z}}$ was estimated as follows \citep[e.g.][]{Dubrulle+1995}
\begin{equation}
    \label{eq:alphaz}
    \alpha_{\mathrm{z}} = \mathrm{St} \; \left[ \left( \frac{H_{\mathrm{g}}}{H_{\mathrm{d}}} \right)^2 - 1 \right]^{-1} \,,
\end{equation}
with $\mathrm{St}$ being the Stokes number, $H_{\mathrm{g}}$ the gas scale height, and $H_{\mathrm{d}}$ a characteristic scale representing the dust vertical extent.
In order to estimate those quantities, the SPH particles were binned in radius. In each radial bin, the Stokes number of the particles was geometrically averaged. To compute $H_{\mathrm{g}}$, we first fitted a Gaussian to the distribution of the particles in altitude for each radial bin. Then, we took $H_{\mathrm{g}}$ as the half-width at half-maximum. $H_{\mathrm{d}}$ was estimated following the same procedure, but by considering only particles with $\delta_{\text{dg}}>0.01$. Figure \ref{fig:alphaz} shows the resulting profiles. Note that the profiles are shown for the radial bins between $a_{\text{cav}}$ and $R_{\mathrm{out}}$ (see Appendix \ref{app : disc size}).

We first see that the turbulence level is, in average, of $\alpha_{\mathrm{z}}\sim5\times10^{-3}$. This is consistent with the input value of $\alpha_{\text{SS}}=5\times10^{-3}$ given as an initial condition of the SPH simulations. At the inner edge of the discs, $\alpha_{\mathrm{z}}$ plummets in the case of circumstellar discs while it increases drastically in the case of circumbinary discs. It highlights that the inner regions of discs around single stars are quiescent locations where the dust is able to settle, reuniting ideal conditions to its growth. Conversely, the inner regions of circumbinary discs are turbulent, which prevents the vertical settling of dust grains and subsequently their growth.

\mdy{}{In circumbinary discs, the turbulence in the inner disc may lead to temporal and azimuthal variations in the gas and dust scale heights. Consequently, it would create vertical motions in the inner disc, that we measured in terms of an effective turbulent diffusion coefficient $\alpha_z$ (see Equation \ref{eq:alphaz}). To investigate variability in the inner circumbinary disc regions, we evolved \textit{IB0} and \textit{IB5} for a short time at high temporal resolution from the timestep used for the main analysis. We integrated the simulations for one orbit of the lump located at the cavity edge. During this orbit, we measured $H_d$ and $H_g$ as a function of both azimuth and time. To compute $H_d$ and $H_g$, we followed the procedure described earlier in this appendix, except that particles were binned in azimuth rather than in radius. In each azimuthal bin and following the results of Appendix \ref{app : disc size}, we considered particles between $5$~au and $20$~au for \textit{IB0}, and between $10$~au and $25$~au for \textit{IB5}. }

\begin{figure*}
\sidecaption
  \includegraphics[width=12cm, trim={0cm 0cm 10cm 0cm}]{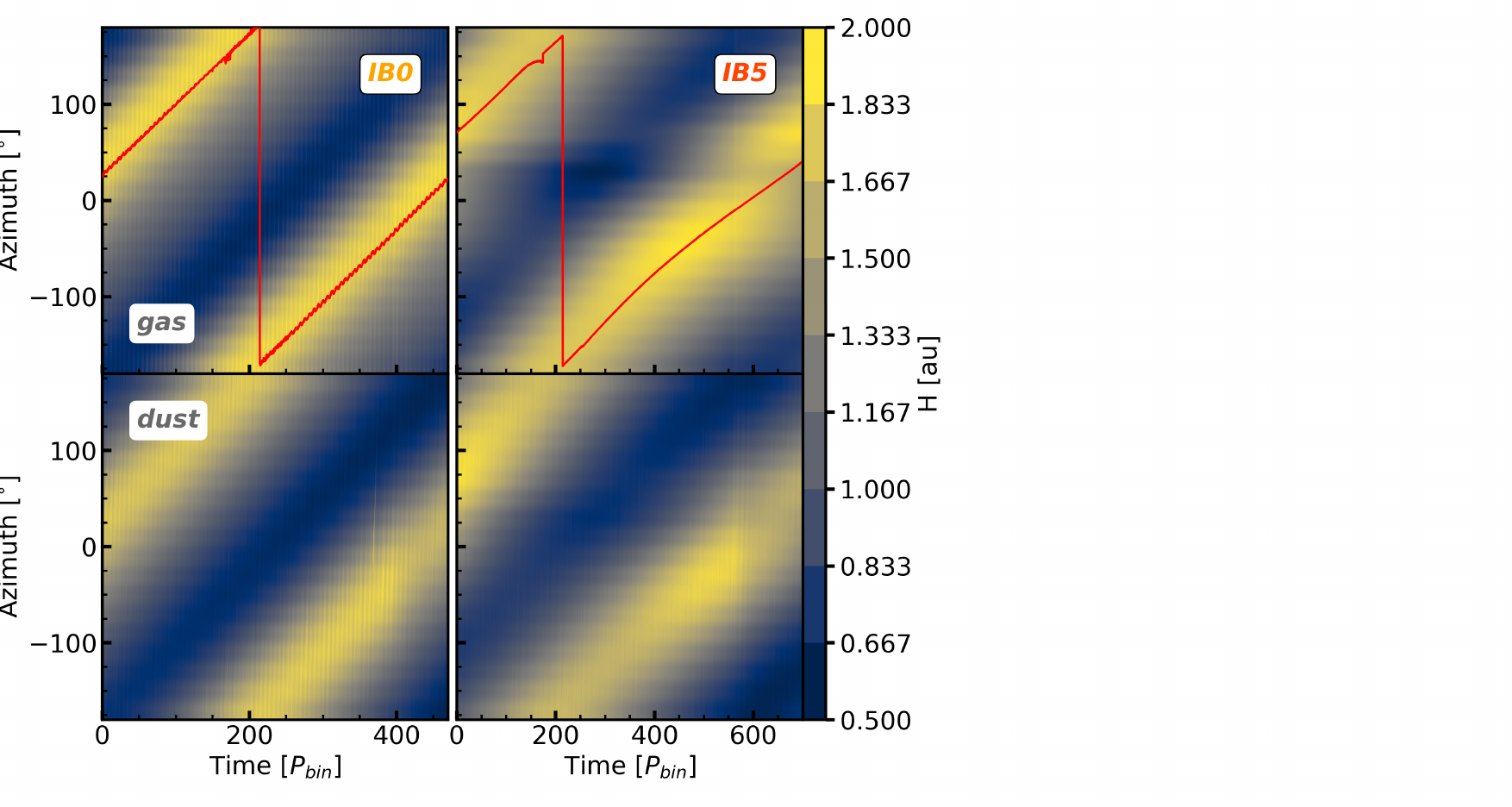}
     \caption{Scale height as a function of time and azimuth. The left column shows \textit{IB0} and the right column shows \textit{IB5}. The top row corresponds to the gas, while the bottom row to the dust. The red curve shows the azimuth of the apocentre of the eccentric cavity as a function of time. To calculate $H$, we considered particles between $5$~au and $20$~au for \textit{IB0}, and between $10$~au and $25$~au for \textit{IB5}.}
     \label{fig:HgHd}
\end{figure*}

\mdy{}{Figure \ref{fig:HgHd} shows $H_d$ and $H_g$ as a function of time and azimuth for one orbit of the lump in \textit{IB0} and \textit{IB5}. We see that both $H_d$ and $H_g$ are modulated as a function of azimuth and time. We observe that $H$ oscillates along the azimuth, with the minimum and maximum values being separated by $180\degree$. The red curve traces the position of the apocentre of the eccentric cavity. Along the orbit of the lump, the maximum of $H$ correlates with the apocentre of the cavity. This result was first characterised by \cite{Ragusa+2024}. They found that gas particles were undergoing vertical oscillations along the course of their orbit. They found that such oscillations were anti-symmetric with respect to the cavity eccentricity vector. As a consequence, the disc vertically compresses towards pericentre, and expands at apocentre. We retrieve this behaviour in \textit{IB0} \textit{IB5}. 
Table \ref{table:HgHd} gives more quantitative details about the observed modulations. Between the maximum and minimum values of $H$, their is a ratio of $\sim2$ in the dust and $\gtrsim1$ in the gas. While the coupled dust is entrained in the oscillations, its absolute $H$ remains small compared to that of the gas. Finally, we compared the orbital timescale of the lump to the orbital time at the gas cavity edge (taken according to Table \ref{table:radii}). We find that the lump orbital timescale, which is similar between gas and dust, is long compared to that at the cavity edge. This confirms that the lump is not produced by a vortex, which would orbit at local Keplerian speed, but is rather a consequence of the non-zero disc eccentricity \citep{KleyDirksen2006,Ataiee+2013,Ragusa+2024}. }

\begin{table*}
    \centering
    \caption{Scale height variations over one orbit of the lump.}
\begin{tabular}{c | c c| c c}
             & \multicolumn{2}{c}{\textit{IB0}} & \multicolumn{2}{c}{\textit{IB5}} \\
             & Dust & Gas & Dust & Gas \\
 \hline\hline
$<h/H>_t$                      & $0.48$ & $0.48$ & $0.19$ & $0.19$ \\
$<H_{\max}/H_{\min}>_t$        & $2.18$ & $1.28$ & $2.49$ & $1.45$ \\
$\min([H_{\max}]_t)$ (au)      & $0.20$ & $1.72$ & $0.14$ & $1.90$ \\
$\max([H_{\max}]_t)$ (au)      & $0.23$ & $1.77$ & $0.17$ & $2.01$ \\
$P_H$ (P$_{\text{bin}}$)  & $475\pm7$ & $480\pm7$ & $656\pm2$ & $641\pm2$ \\
$P_H$ (P$_{\text{cav}}$)  & $258\pm4$ & $261\pm4$ & $183\pm1$ & $179\pm1$
\end{tabular}
\tablefoot{$<h/H>_t$ is the smoothing length averaged over time normalised by $H$ at $a_{\text{cav}}$. $<H_{\max}/H_{\min}>_t$ is the ratio of the maximum and minimum $H$ along the azimuth averaged over time. Considering the maximum of $H$ at each timestep, $\min([H_{\max}]_t)$ (resp. $\max([H_{\max}]_t)$) is the minimum (resp. maximum) of these values. $P_H$ is the azimuthal period of the lump, measured by fitting the position of the maximum $H$ along the azimuth at each timestep.}
\label{table:HgHd}
\end{table*}

\section{Kepler and directly imaged circumbinary planets}
\label{app : cbp}

\mdytwo{}{Tables \ref{table:kepler_planets} and \ref{table:DI_planets} list the properties of the circumbinary planets detected by Kepler and by direct imaging, respectively.}

\begin{table*}[h!]
\caption{Characteristics of confirmed transiting \textit{Kepler} circumbinary planets.} 
\centering
 \begin{tabular}{c c c c c} 
 Name & Planet separation (a$_{\text{bin}}$) & Planet eccentricity & Planet mass (M$_{\text{J}}$) & References\\
 \hline \hline

Kepler-47 & $3.53\pm0.02$ & $0.021\pm0.002$ & $0.007\pm0.075$ & \cite{Orosz+2012,Orosz+2019}\\
          & $8.58\pm0.06$ & $0.024\pm0.025$ & $0.060\pm0.075$ & \cite{Orosz+2012,Orosz+2019}\\
          & $11.83\pm0.08$ & $0.044\pm0.029$ & $0.010\pm0.007$ & \cite{Orosz+2019}\\

Kepler-413 & $3.50\pm0.028$ & $0.118\pm0.113$ & $0.210\pm0.067$ & \cite{Kostov+2014} \\

Kepler-1647 & $21.32\pm0.06$ & $0.058\pm0.069$ & $1.520\pm0.648$ & \cite{Kostov+2016} \\

Kepler-38 & $3.16\pm0.09$ & $<0.032$ & $<0.384$ & \cite{Orosz+2012-38} \\

Kepler-35 & $3.43\pm0.01$ & $0.042\pm0.007$ & $0.127\pm0.021$ & \cite{Welsh+2012} \\

Kepler-64 & $3.64\pm0.09$ & $0.054\pm0.008$ & $<0.532$ & \cite{Schwamb+2013} \\

Kepler-453 & $4.26\pm0.07$ & $0.036\pm0.009$ & $0.001\pm0.050$ & \cite{Welsh+2015} \\

Kepler-1661 & $3.39\pm0.05$ & $0.057\pm0.005$ & $0.053\pm0.038$ & \cite{Socia+2020} \\

Kepler-16 & $3.14\pm0.01$ & $0.007\pm0.001$ & $0.333\pm0.016$ & \cite{Doyle+2011} \\

Kepler-34 & $4.76\pm0.01$ & $0.182\pm0.020$ & $0.220\pm0.011$ & \cite{Welsh+2012} \\
 \hline

Average & $4.97\pm0.02$ & $0.043\pm0.008$ & $0.058\pm0.467$ &\\
\end{tabular}
\tablefoot{The \textit{Average} row lists the column-wise geometric average. Upper limits were not taken into account in the calculation of the averages.}
\label{table:kepler_planets}
\end{table*}

\begin{table*}[h!]
\caption{Characteristics of directly imaged circumbinary planets.} 
\centering
 \begin{tabular}{c c c c} 
 System name & Planet separation\tablefootmark{1} (a$_{\text{bin}}$) & Planet mass (M$_{\text{J}}$) & References\\
 \hline \hline
SR12 & $42\pm4$ & $14\pm7$ & \cite{Kuzuhara+2011}\tablefootmark{2}\\
 
Delorme AB &$7\pm1$ & $13\pm1$ & \cite{Delorme+2013}\\

M6 VL-G & $899\pm130$ & $12\pm1$ & \cite{Dupuy+2018}\\

b~Cen & Unknown & $11\pm2$ & \cite{Janson+2021}\\

VHS J1256-1257 & $179\pm77$ & $12\pm1$ & \cite{Dupuy+2023}\tablefootmark{3}\\\

HD~106906 & $5230\pm22$ & $11\pm2$ & \cite{Bailey+2014,Rouan+2025}\\

WISPIT~1 & $32\pm2$ & $10\pm1$ & \cite{vanCapelleveen+2025}\\
         & $80\pm4$ & $5\pm1$ & \cite{vanCapelleveen+2025}\\

HD~143811 & $61\pm9$ & $6\pm1$ & \cite{Squicciarini+2025}\\
\hline
Average & $118\pm8$ & $10\pm1$ & 
\end{tabular}
\tablefoot{The \textit{Average} row lists the column-wise geometric average.\\
\tablefootmark{1}{When not available, $a_{\text{bin}}$ was taken as the projected separation between the binary components. If the projected separation is unknown as well, we label the planet separation as \textit{Unknown}.\\
\tablefootmark{2}{For SR12 and due to the lack of a precise measurement, the error on the planet separation was assumed to be $10\%$.}\\
\tablefootmark{3}{For VHS J1256-1257, we note that another model in the same reference finds a planet mass of $16\pm1$~M$_{\text{J}}$.}
}
}
\label{table:DI_planets}
\end{table*}


\end{appendix}

\end{document}